\documentclass[11pt,letterpaper]{article}
\pdfoutput=1
\usepackage{jheppub}
\usepackage{bbm}
\usepackage{mathrsfs}
\usepackage{slashed}
\usepackage{caption}
\usepackage{epstopdf}
\usepackage[normalem]{ulem}
\usepackage[bottom]{footmisc}
\usepackage{subcaption}
\usepackage{bbold}
\usepackage{titlesec}
\usepackage{threeparttable}
\usepackage{booktabs}
\usepackage{changepage}
\usepackage[utf8]{inputenc}
\usepackage{dsfont}

\usepackage{grffile}
 
\usepackage{graphicx}  
\usepackage{dcolumn}   
\usepackage{bm}        
\usepackage{amssymb}   
\usepackage{setspace}
\usepackage{amsmath, amssymb, setspace}
\usepackage{array}
\usepackage{booktabs}
\usepackage{caption}
\usepackage{indentfirst}
\usepackage{float}
\usepackage{lmodern}
\usepackage{multirow}
\usepackage{soul}
\usepackage[normalem]{ulem}

\usepackage{braket}
\usepackage{comment}

\usepackage[draft]{pgf}

\usepackage{adjustbox} 
\newlength{\myimageoversize}
\newsavebox{\myimage}
\newcommand{\mycenter}[1]{%
\savebox{\myimage}{#1}
\settowidth{\myimageoversize}{\usebox{\myimage}}
\addtolength{\myimageoversize}{-\textwidth}
\setlength{\leftskip}{-0.5\myimageoversize}
\noindent
\usebox{\myimage}}

\usepackage[most]{tcolorbox}

\usepackage{empheq}

\tcbset{colback=white!10!white, colframe=black!50!black, 
        highlight math style= {enhanced, 
            colframe=black,colback=white!10!white,boxsep=0pt}
        }

\titleformat{\subsection}
 {\normalfont\fontsize{12}{17}\itshape}{\thesubsubsection}{1em}{}
 
\title{\huge{Cosmological Dependence of Resonantly Produced Sterile Neutrinos}}

\author[a]{Graciela B. Gelmini,}
\author[a]{Philip Lu}
\author[a]{and Volodymyr Takhistov}

\affiliation[a]{Department of Physics and Astronomy, University of California, Los Angeles\\
 Los Angeles, CA 90095-1547, USA}

\emailAdd{gelmini@physics.ucla.edu}
\emailAdd{philiplu11@gmail.com}
\emailAdd{vtakhist@physics.ucla.edu}

\abstract{
The detection of a sterile neutrino could constitute the first observation of a particle that could have been produced before Big-Bang Nucleosynthesis (BBN), and could provide information about the yet untested pre-BBN era. The cosmological evolution in this era could be drastically different than typically assumed in what constitutes the standard cosmology, as happens in a variety of motivated particle  models. In this work we assess the sensitivity to different pre-BBN cosmologies in which entropy is conserved of 0.01~eV to 1~MeV mass sterile neutrinos produced in the early Universe via resonant active-sterile oscillations, which  requires a large lepton asymmetry. We identify mass ranges where it is possible to have two populations of the same sterile neutrino, one with a colder and one with a hotter momentum spectra, which is in principle an observable effect. Furthermore, we show the regions in mass and mixing where fully resonant production (i.e.~simultaneously coherent and adiabatic) can occur.  We find that in several of the cosmologies we consider, including the standard one, for a lepton asymmetry larger than $\sim10^{-4}$  fully resonantly produced sterile neutrinos in the eV-mass range can evade all cosmological constraints. 
}

\begin{document}
\preprint{}
 \maketitle
\flushbottom

\section{Introduction}

Light nuclei produced during the Big Bang Nucleosynthesis (BBN) are the earliest cosmological remnants that have been detected thus far. Hence, the cosmological history of the Universe prior to BBN that occurred at temperatures $T>$ 5 MeV has not been tested and remains unknown~\cite{deSalas:2015glj, Hasegawa:2019jsa, DeBernardis:2008zz, Hannestad:2004px, Kawasaki:2000en, Kawasaki:1999na}. Since many dark matter (DM) particle candidates are produced during this unknown era, assumptions must be made to compute their relic density and momentum distribution. Typically, it is considered that the cosmological history is a simple extension to higher temperatures (earlier times) of the late cosmology we know at temperatures below $T\sim 5$ MeV. In particular, it is assumed that the Universe was radiation-dominated, that only the Standard Model (SM) particle content was present and that there was no  extra entropy change in matter and radiation. These assumptions constitute the ``standard pre-BBN cosmology''. However, in many motivated particle models, such as those based on moduli decay, quintessence or extra dimensions, the cosmological history could be  different. This could drastically affect the resulting relic density and momentum distribution of particles
produced before BBN. Thus detecting any such particle may provide invaluable information about this yet uncharted epoch.

A promising relic to consider for testing the early Universe cosmology is a sterile neutrino. Sterile neutrinos appear in very well motivated extensions of the SM. The SM contains three massless ``active" neutrinos $\nu_{\alpha}$, characterized by their $\alpha = e, \mu, \tau$ flavor, which interact weakly. The discovery of neutrino oscillations~\cite{Fukuda:1998mi} proved that neutrinos are massive and  the inclusion of ``sterile'' neutrino species $\nu_s$ is one of the simplest possibilities to give active neutrinos a non-zero mass. In the simplest models, sterile neutrinos only mix with SM particles via their mixings with active neutrinos. This is what we assume in this work. For simplicity, we assume a $\nu_s$ that mixes with only one of the $\nu_{\alpha}$, which for our figures is $\nu_e$, with a mixing angle of $\sin \theta$.

A broad range of possible sterile neutrino masses has been discussed in the literature. While the three-flavor paradigm of active neutrino oscillations has been extensively tested~\cite{Cleveland:1998nv,Abdurashitov:2009tn,Altmann:2005ix,Hampel:1998xg,Aharmim:2005gt,Abe:2016nxk,Araki:2004mb,An:2012eh,Abe:2013sxa,Ahn:2012nd,Ahn:2006zza,Adamson:2013whj,Abe:2014ugx,Agafonova:2014bcr,Adamson:2016xxw,Aartsen:2014yll,Fukuda:1998mi}, several laboratory anomalies\footnote{We stress that such signals are in strong tension with other results, e.g. from IceCube~\cite{TheIceCube:2016oqi} and MINOS~\cite{Adamson:2017uda}.} (e.g. LSND~\cite{Aguilar:2001ty} and MiniBooNE~\cite{Aguilar-Arevalo:2013pmq,Aguilar-Arevalo:2018gpe} as well as DANSS~\cite{Alekseev:2018efk} and NEOS~\cite{Ko:2016owz}) appear to be consistent with additional sterile neutrinos of $m_s = \mathcal{O}$(eV) mass. Reactor neutrino data~\cite{An:2016luf,Declais:1994su,Ashenfelter:2018iov} constrain these sterile neutrinos and future data from the PTOLEMY~\cite{Betti:2019ouf} and KATRIN~\cite{megas:thesis}  experiments will further test them.

Sterile neutrinos with mass of $\mathcal{O}$(keV) and a spectrum close to thermal constitute a viable Warm DM (WDM) candidate (see e.g. Ref.~\cite{Boyarsky:2018tvu}).
Strong astrophysical bounds (e.g. from Lyman-$\alpha$ forest and X-ray data) disfavor them as the dominant DM component if they are produced through
active-sterile oscillations in the early Universe~(e.g.~\cite{Palazzo:2007gz}). It has been also suggested that a 3.5-keV X-ray emission line signal from galaxy clusters can be attributed to sterile neutrinos of $m_s = 7$ keV~\cite{Bulbul:2014sua,Boyarsky:2014jta}. Sterile neutrinos of $m_s \gtrsim$ keV could play a significant role in supernovae and explain the
observed pulsar velocities~\cite{Fuller:2003gy}.
Future data from KATRIN/TRISTAN~\cite{Mertens:2018vuu} and HUNTER~\cite{Smith:2016vku} experiments will provide a stringent test of $\mathcal{O}$(keV) mass sterile neutrinos. Even heavier sterile neutrinos with mass $m_s \gtrsim \mathcal{O}(10)$ MeV have been recently proposed to alleviate the discrepancy between local and cosmological measurements of the Hubble constant~\cite{Gelmini:2019deq}.

In the early Universe, sterile neutrinos without additional interactions beyond the SM  that couple to the SM particles only through mixing with active neutrinos,  as we assume here, are produced through active-sterile  flavor oscillations and collisional processes. These oscillations can be resonant or non-resonant. In non-minimal particle models, sterile neutrino production could also proceed via other mechanisms, e.g. in the decay of additional heavy scalars~\cite{Petraki:2007gq}. 

In the absence of a large lepton asymmetry the oscillations are non-resonant and sterile neutrinos are produced via the Dodelson-Widrow (DW) mechanism~\cite{Dodelson:1993je}. In the standard cosmology this mechanism results in a Fermi-Dirac relic momentum distribution of sterile neutrinos, with a reduced magnitude with respect to active neutrinos. Several studies have already been carried out on non-resonantly produced sterile neutrinos in different pre-BBN cosmologies 
(e.g. in Refs.~\cite{Gelmini:2004ah, Gelmini:2008fq, Rehagen:2014vna,Abazajian:2017tcc, Gelmini:2019esj, Gelmini:2019wfp}). This work is complementary to these previous studies in that we concentrate on the production of sterile neutrinos via resonant active-sterile oscillations,  an often considered production mechanism that requires a significant lepton asymmetry in active neutrinos (this is the Shi-Fuller mechanism~\cite{Shi:1998km},  see also Ref.~\cite{Abazajian:2001nj,Abazajian:2004aj}). In this case, sterile neutrinos are produced with a  colder momentum distribution  (i.e. with a lower average momentum) that is different from a Fermi-Dirac spectrum, even in the standard cosmology.  In particular,  we study the effect of different cosmologies on resonantly produced sterile neutrinos with mass $10^{-2}$ eV $< m_s <$ 1 MeV. Since the production rate is usually not fast enough for sterile neutrinos to equilibrate, the final relic abundance and spectrum are fixed by freeze-in.
  
We will analyze resonant sterile neutrino production within several example cosmological models in which entropy is conserved, characterized by the magnitude and temperature dependence of the Hubble expansion rate $H$ in the non-standard cosmological phase. If $H$ is larger than it would be in the standard cosmology, the resonant production of sterile neutrinos during this phase is suppressed with respect to standard production, and if $H$ is lower, the production is enhanced. If the production is fully resonant, which requires adiabaticity and coherence at the resonance, the relic sterile neutrino number density depends only on the lepton number and not on the expansion rate. However,  the mass-mixing regions where this type of production occurs move to smaller (larger) mixing angles for smaller (larger) $H$ values with respect to what it would be in the standard cosmology. Though it 
is difficult to achieve in any non-standard cosmology an expansion rate that is smaller than it would be in the standard cosmology, it is possible, e.g. in a 
particular scalar-tensor model. 
 
This paper is organized as follows. In Section~\ref{sec:modcos} we describe a
parametrization of the expansion rate of the Universe $H$ and define the four particular pre-BBN cosmological models that we consider. In Sec.~\ref{sec:res} we discuss resonant production in general, and fully resonant production in different  pre-BBN cosmologies. In Sec.~\ref{sec:limits} we briefly describe sterile neutrino constraints and regions of interest in the mass-mixing plane, in Sec.~\ref{sec:mainresults} we summarize our main results and in Sec.~\ref{sec:summary} we conclude.
 
\section{Early Universe cosmology}
\label{sec:modcos}

The expansion rate of the Universe $H(T)$ is determined by the Friedmann equation. 
The standard cosmological model~\cite{Kolb:1990vq} assumes that before BBN the Universe was radiation-dominated, with the temperature $T$  of the radiation bath reaching values much larger than the temperature at which BBN starts (i.e. $T\sim$ MeV). In this case $H$ is
\begin{equation} 
\label{eq:hStd}
H_{\rm Std} =  \Big(\dfrac{T^2}{M_{\rm Pl}}\Big)  \sqrt{\dfrac{8 \pi^3 g_{\ast}(T)}{90}}~.
\end{equation}
Here $M_{\rm Pl}= 1.22 \times 10^{19}$ GeV is the Planck mass and $g_\ast(T)$ is the number of degrees of freedom contributing to the energy density at temperature $T$ (see e.g.~Ref.~\cite{Husdal:2016haj,Borsanyi:2016ksw,Drees:2015exa}). With only the SM degrees of freedom present, $g_\ast=80$ above the QCD phase transition that takes place at $T \simeq 200$ MeV. It decreases steeply close to the QCD phase transition, with a characteristic value of $g_{\ast} \simeq 30$ until $T$ decreases to $20$ MeV. Between this temperature and $T =1$ MeV, when electrons and positrons become non-relativistic and annihilate, $g_{\ast} = 10.75$. For simplicity, in our figures we use  $g_\ast=10.75$.

In the non-standard cosmologies we consider, the cosmological evolution of the early Universe is allowed to be  drastically different as long as the cosmology reduces to the standard one at $T < 5$ MeV, in order to not spoil the known late history~\cite{deSalas:2015glj, Hasegawa:2019jsa, DeBernardis:2008zz, Hannestad:2004px, Kawasaki:2000en, Kawasaki:1999na}.
We explore here non-standard cosmologies where entropy in matter and radiation is conserved and the expansion rate in the non-standard phase can be described by a  phenomenological parameterization in terms of three real parameters, $\eta$,  $\beta$ and $T_{\rm tr}$, which captures a large class of models~\cite{Catena:2009tm} 
\begin{equation} \label{eq:hnStd}
    H = \eta~ \Big(\dfrac{T}{T_{\rm tr}}\Big)^{\beta} H_{\rm Std}~.
\end{equation}
Here $T_{\rm tr}$ is the transition temperature  below which the cosmology becomes standard, and $\eta > 0$.
Following our previous study of non-resonant sterile neutrino production~\cite{Gelmini:2019esj, Gelmini:2019wfp}, we consider several specific examples of such cosmologies:
\begin{itemize}
\item kination (K)~\cite{Spokoiny:1993kt,Joyce:1996cp,Salati:2002md,Profumo:2003hq,Pallis:2005hm} (with $\eta = 1$, $\beta = 1$)
\begin{align} 
\label{eq:hkin}
H_{\rm K}= \Big(\dfrac{T^3}{M_{\rm Pl} T_{\rm tr}}\Big) \sqrt{\dfrac{8\pi^3 g_{\ast}}{90}}
= \Big(\dfrac{T}{T_{\rm tr}}\Big) ~ H_{\rm Std} 
\end{align}
\item scalar-tensor model (ST1) of Ref.~\cite{Catena:2004ba}  (with $\eta = 7.4 \times 10^5$, $\beta = -0.8$)
\begin{equation} \label{eq:hst1}
H_{\rm ST1} = 7.4 \times 10^5 \Big(\dfrac{ T_{\rm tr}^{0.8}~T^{1.2}}{M_{\rm Pl}}\Big)   \sqrt{\dfrac{8\pi^3 g_{\ast}}{90}} = 7.4 \times 10^5 \Big(\dfrac{T}{T_{\rm tr}}\Big)^{-0.8} ~ H_{\rm Std} 
\end{equation}
\item scalar-tensor model (ST2) of Ref.~\cite{Catena:2007ix}  (with $\eta = 0.03$, $\beta = 0$)
\begin{equation} 
\label{eq:hst2}
H_{\rm ST2} = 3.2 \times 10^{-2} \Big(\dfrac{1}{M_{\rm Pl}}\Big) T^{2}  \sqrt{\dfrac{8\pi^3 g_{\ast}}{90}}= 0.03 ~ H_{\rm Std}
\end{equation}
\end{itemize}
Our analysis readily extends to scalar-tensor models whose expansion rate is in between $H_{\rm ST1}$ and $H_{\rm ST2}$~\cite{Catena:2004ba, Catena:2007ix}.
We choose $T_{\rm tr} = 5$~MeV throughout, ensuring consistency with BBN.

Further discussion of these cosmologies 
can be found in our previous study of non-resonant sterile neutrino production and references therein~\cite{Gelmini:2019wfp}.
Since we impose that any non-standard cosmological evolution transitions to the standard cosmology prior to BBN, there are no additional constraints from astrophysical observations (e.g. binary star mergers~\cite{Sakstein:2017xjx}).

\section{Resonant sterile neutrino production}
\label{sec:res}
\subsection{Boltzmann equation}
\label{ssec:boltzmann}

As in the case of non-resonant sterile neutrino production~\cite{Gelmini:2019wfp}, for resonant production the time evolution of the momentum
distribution function of sterile neutrinos $f_{\nu_s}(p,t)$ is given by 
the Boltzmann equation
~\cite{Kolb:1990vq,Abazajian:2001nj, Rehagen:2014vna}
\begin{equation}
\label{eq:boltzmann2}
    -HT\left(\frac{\partial f_{\nu_s}(E,T)}{\partial T}\right)_{E/T=\epsilon} \simeq~ \Gamma(E,T)f_{\nu_\alpha}(E,T)~,
\end{equation}
where $f_{\nu_{\alpha}}(p,t) = (e^{\epsilon-\xi}+1)^{-1}$ is the density function of active neutrinos, $p$ is the magnitude of the neutrino momentum, $\epsilon = p/T$ is the dimensionless momentum,  $\xi = \mu_{\nu_{\alpha}}/T$ is the $T$-scaled dimensionless chemical potential where $\mu_{\nu_{\alpha}}$ is the chemical potential of $\nu_{\alpha}$,  $\Gamma (p,t)$ is the conversion rate of active to sterile neutrinos, the energy is $E \simeq p$, since all neutrinos of interest are relativistic at production, and derivative in the left-hand side of the equation is computed at constant $\epsilon$. Eq.~\eqref{eq:boltzmann2} is valid if
 $f_{\nu_s} \ll 1$ and $f_{\nu_s} \ll f_{\nu_{\alpha}}$.

The total rate $\Gamma$ of active to sterile  neutrino conversion is given by the probability $\langle P_m \rangle$ of an active-sterile flavor oscillation in matter (for  mixing angle in matter $\theta_m$) times the interaction rate
\begin{equation} 
\label{eq:interaction-active}
    \Gamma_{\alpha} = d_{\alpha} G_F^2 \epsilon T^5,
\end{equation}
 of active neutrinos with the surrounding plasma~\cite{Abazajian:2001nj}
\begin{equation} 
\label{eq:interaction}
    \Gamma ~=~ \dfrac{1}{2}  \langle P_m (\nu_\alpha \rightarrow \nu_s) \rangle \Gamma_{\alpha}   ~\simeq~ \frac{1}{4}\sin^2(2\theta_m)d_\alpha G_F^2\epsilon T^5~,
\end{equation}
where $d_{\alpha} = 1.27$ for $\nu_e$ and $d_\alpha = 0.92$ for $\nu_{\mu}$, $\nu_{\tau}$. 
The mixing angle  in matter $\theta_m$ is given by
\begin{equation}
\label{eq:mattermixing}
    \sin^2(2\theta_m) = \frac{\sin^2(2\theta)}{\sin^2(2\theta) + \Big[\cos(2\theta)-2\epsilon T (V_D+V_T)/m_s^2\Big]^2}~.
\end{equation}
In the right hand side of Eq.~\eqref{eq:boltzmann2} we have omitted the quantum damping factor $[1- (\Gamma_\alpha \ell_m/2)^2]^{-1}$ where $\ell_m$ is the oscillation length in matter (see e.g. the discussion in Refs.~\cite{Abazajian:2001nj,Kishimoto:2008ic}). This term is typically negligible for the range of parameters relevant for our study, but could become significant at the resonance (because at resonance the neutrino oscillation length is maximal),  as we discuss in Sec.~\ref{ssec:coherenceadiabaticity}. 

In the denominator of Eq.~\eqref{eq:mattermixing} $V_T$ is the thermal potential
\begin{align}
 V_T ~&=~ -B\epsilon T^5~, 
 \end{align}
where $B$ is a constant prefactor dependent on flavor (given below in parenthesis) and temperature range  (given to the right of each line below)
\begin{align} \label{eq:bprefac}
\begin{split}
    B= \left\{
                \begin{array}{llc}
                  10.88\times10^{-9}~\textrm{GeV}^{-4}~(e);
                  &3.02\times10^{-9}~\textrm{GeV}^{-4}~(\mu, \tau);   &T\lesssim20\textrm{ MeV}~\\
                  10.88\times10^{-9}~\textrm{GeV}^{-4}~(e, \mu);  &3.02\times10^{-9}~\textrm{GeV}^{-4}~(\tau); ~~ 20\textrm{ MeV} \lesssim &T \lesssim 180 \textrm{ MeV} \\
                  10.88 \times10^{-9}~\textrm{GeV}^{-4}~(e,\mu,\tau);  & &T\gtrsim180 \textrm{ MeV}
                \end{array}
              \right.
\end{split}
\end{align} 
The density potential  $V_D$
depends on the particle-antiparticle asymmetries in the background  
\begin{align} 
 V_D ~&=~ \frac{2\sqrt{2}\zeta(3)}{\pi^2}G_F T^3 \left(\mathcal{L}\pm\frac{\eta}{4}\right) \simeq 0.34 \,  G_F T^3 \mathcal{L}~,
\end{align}
where $\zeta(3) \simeq
1.202$ is the Riemann zeta function,  $\eta \simeq 6 \times 10^{-10}$ is the baryon-to-photon ratio~\cite{Tanabashi:2018oca} that represents the lepton asymmetry stored in electrons (which is equal to the baryon asymmetry due to charge neutrality), the ``+'' sign is taken for $\alpha = e$ while ``-''  is for $\alpha = \mu, \tau$ and $\mathcal{L}$ is the lepton number\footnote{This is also known as ``potential lepton number''~\cite{Kishimoto:2008ic}.} in active  neutrinos,
\begin{equation}
\mathcal{L} = 2 L_{\nu_\alpha} + \sum\limits_{\beta\neq\alpha}L_{\nu_\beta}~.
\end{equation}
The individual lepton number for each active neutrino flavor is $L_{\nu_{\alpha}} = (n_{\nu_\alpha} - n_{\overline{\nu}_\alpha})/n_\gamma$,  where $n_{\gamma}$ is the photon number density. 
BBN constrains the magnitude of the electron neutrino asymmetry to be $|L_{\nu_e}|\lesssim \mathcal{O}(10^{-3})$~\cite{Mangano:2011ip}. The upper limit imposed by BBN and CMB on the asymmetries in the other neutrino flavors is larger,  $|L_{\nu_{\mu}, \nu_{\tau}}| \lesssim \mathcal{O}(10^{-1})$~\cite{Barenboim:2016shh}.  Throughout our discussion $\mathcal{L} \gg \eta$ and hence $\eta$ is neglected. If neutrino oscillations efficiently redistribute the lepton asymmetry among the flavors, the resulting lepton number is
 \begin{equation}
 \label{eq:3/4L}
  \mathcal{L} \simeq \frac{4}{3} \sum\limits_{\alpha} L_{\nu_\alpha}~.
\end{equation}

For non-resonant (DW) sterile neutrino production, for which $V_D$ is negligible, the maximum of the momentum-integrated production rate over $H$ (namely $(\gamma/H)=d  (n_{\nu_s}/n_{\nu_\alpha})/d \ln{T}$, as considered in the DW paper ~\cite{Dodelson:1993je} - see App.~\ref{ssecapp:maxtemp}) occurs at the temperature $T_{\rm max}$, which for the standard cosmology is~\cite{Dodelson:1993je,Kainulainen:1990ds}
\begin{equation}
 \label{eq:Tmaxstd}
  T_{\rm max}^{\rm Std} \simeq 108 ~ {\rm MeV}~\Big(\dfrac{m_s}{\rm keV}\Big)^{1/3}~. 
\end{equation}
The temperature $T_{\rm max}$ does not change significantly in the different cosmologies we consider  (see App.~\ref{ssecapp:maxtemp}).

\subsection{Resonance conditions}
\label{rescondition}

For resonant sterile neutrino production~\cite{Shi:1998km} we follow the discussion of Ref.~\cite{Abazajian:2001nj}. A resonant neutrino conversion occurs when the mixing angle in matter $\sin^2(2\theta_m)$ is maximized. Thus, the condition for resonance is that the square bracket in the denominator of Eq.~\eqref{eq:mattermixing}, denoted here as $R(T)$, vanishes (i.e. $\sin^2(2\theta_m)= 1$) 
\begin{equation}
\label{eq:resonance}
    R(T) \equiv m_s^2 \cos(2\theta) -2\epsilon T (V_D +   V_T) \simeq m_s^2 - aT^4 + bT^6 = 0~.
\end{equation}
Here,  $a \simeq 0.69 \, G_F \epsilon  \mathcal{L}$, $b=2 \epsilon^2 B$ and $\cos(2 \theta) \simeq 1$ for all the mixing angles we consider. In general there are two roots of this equation. Since the lower temperature root corresponds to a larger production rate, it is the only one we consider.
  In the presence of a sizable lepton asymmetry
  the dominant resonance occurs at a temperature~\cite{Shi:1998km}
  \begin{equation}
\label{eq:tres}
    T_\textrm{res} = 18.8~ \textrm{MeV}~  \epsilon^{-\frac{1}{4}}\left(\frac{m_s}{\textrm{keV}}\right)^{\frac{1}{2}}\mathcal{L}^{-\frac{1}{4}}~.
\end{equation}
Due to the inverse dependence of $T_\textrm{res}$ on $\epsilon = p/T$, active neutrinos with lower momentum undergo resonant conversion earlier, at higher temperatures. Thus, with increasing time (decreasing temperature) the resonance ``sweeps" from the low momentum end of the active neutrino distribution towards higher momenta, converting active  to sterile neutrinos and reducing $\mathcal{L}$, which in turn further increases the sweep rate $\dot{\epsilon}$ (see Eq. (7.11) of Ref.~\cite{Abazajian:2001nj}).

\begin{figure}
\begin{center}
\includegraphics[scale=.38]{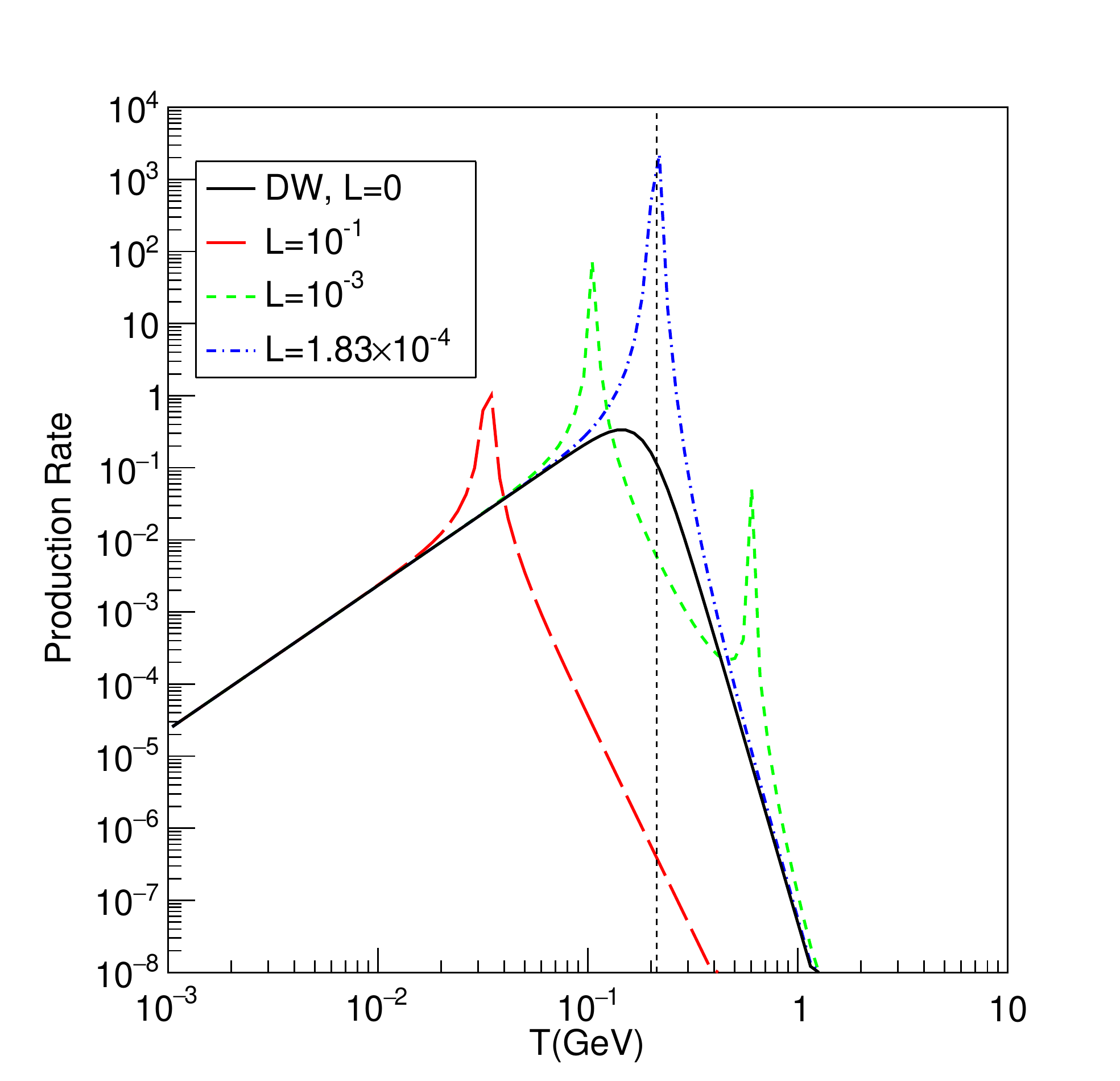}
\includegraphics[scale=.38]{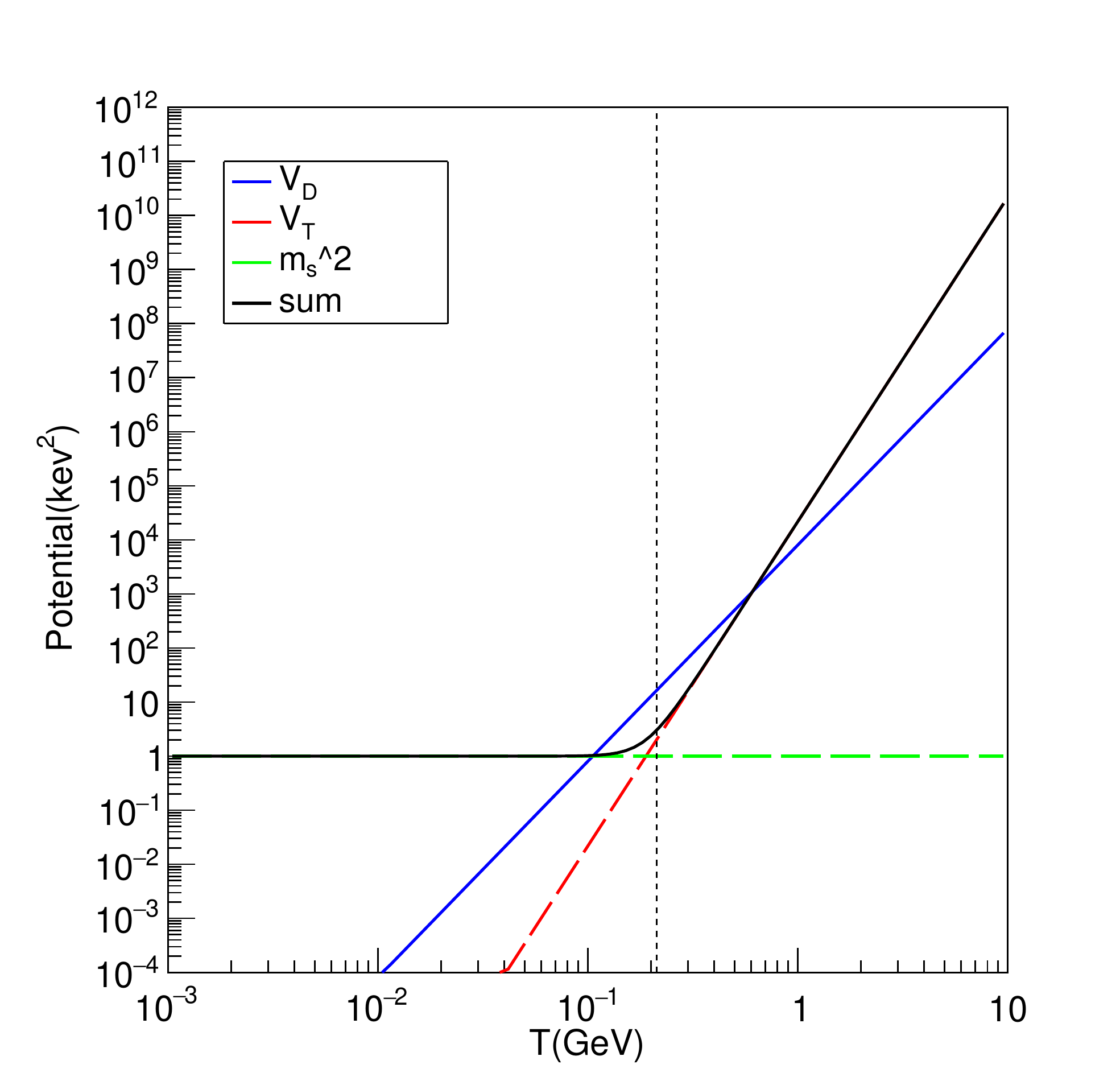}
\caption{
{\small [Left] Production rate ($\partial f_{\nu_s}/\partial T)_{\epsilon}$ for $\epsilon = 1$ and $m_s = 1$ keV in the standard cosmology for non-resonant oscillations with $\mathcal{L}=0$,  i.e.
DW production (black solid line), and resonant oscillations with $\mathcal{L}=10^{-1}$ (red long dashed line), $\mathcal{L}=10^{-3}$ (green short dashed line) and $\mathcal{L}_\textrm{reslim}=1.83\times10^{-4}$ (blue 
dot-dashed line). The peaks  indicate occurrence of resonances. For large $\mathcal{L}$ only the low temperature peak is visible. As $\mathcal{L}$ decreases, the two visible resonant peaks move towards each other, eventually merging into one peak at the critical lepton number  $\mathcal{L}_{\rm reslim}$ of Eq.~\eqref{eq:Lreslimit} below which there is no resonance.
The resonance temperature for $\mathcal{L}_{\rm reslim}$ is denoted by the black dotted line. Notice that the high temperature resonance is always significantly suppressed relative to the low temperature resonance. [Right] Contributions of different terms to the denominator of Eq.~\eqref{eq:mattermixing}, including sterile neutrino mass squared $m_s^2 = 1$ keV$^2$ (green dashed line), thermal potential $V_T$ (red dashed line), density potential $V_D$ (blue solid line) for $\mathcal{L}=10^{-3}$, and sum of the mass and thermal potential terms (black solid line), for $\epsilon = 1$. Resonance occurs when the $V_D$ term (blue line) crosses the sum of $V_T$ and the mass term (black line), corresponding to locations of the peaks for $\mathcal{L}=10^{-3}$ } (green short dashed line) shown in the left panel. For $\mathcal{L} = \mathcal{L}_{\rm reslim}$ the blue and black lines would intersect at only one point and would not cross for $\mathcal{L}<\mathcal{L}_\textrm{reslim}$.} 
\label{fig:restransitions}
\end{center}
\end{figure}

The minimum of $R(T)$,  where $dR/dT= 0$, occurs at a temperature called  $T = T_{\rm PEAK}$ in Ref.~\cite{Abazajian:2001nj}. Imposing that $R(T_{\rm PEAK}) = 0$, which amounts to requiring that the two roots of $R(T)$ converge into one, defines the critical lepton number - which we call $\mathcal{L}_{\textrm{reslim}}$  - below which there are no resonances for a given mass $m_s$ and momentum $\epsilon$,  
\begin{equation}
\label{eq:Lreslimit}
    \mathcal{L}_{\textrm{reslim}} = 1.83\times10^{-4}\epsilon^{\frac{1}{3}} \left(\frac{m_s}{\textrm{keV}}\right)^{\frac{2}{3}}\left(\frac{B}{10.88\times10^{-9}\textrm{ GeV}^{-4}}\right)^{\frac{2}{3}}~.
\end{equation}
Namely, given a sterile neutrino mass $m_s$ and momentum $\epsilon$, resonant production occurs only for $\mathcal{L} > \mathcal{L}_{\rm reslim}$. 

We display a typical resonance behavior in Fig.~\ref{fig:restransitions}. The right panel shows the contributions of the potentials $V_T$ and $V_D$ as well as the mass term to the denominator of Eq.~\eqref{eq:mattermixing}, for $m_s=$ 1 keV and $\mathcal {L}= 10^{-3}$ as function of the temperature. Resonances occur when the $V_D$ term (blue line) crosses the sum of $V_T$ and the mass term (black line). The left panel   shows the production rate $(\partial f_{\nu_s}/\partial T)_{\epsilon}$ as a function of the temperature $T$ for $m_s = 1$ keV, $\epsilon = 1$ and different values of $\mathcal{L}$: $0$ (black line), $10^{-1}$ (long dashed red line), $10^{-3}$ (short dashed green line) and $\mathcal{L}_{\rm reslim}$ (dot-dashed blue line). As $\mathcal{L}$ decreases, the two visible resonant peaks move towards each other, eventually merging into one peak at the critical lepton number  $\mathcal{L}_{\rm reslim}$ of Eq.~\eqref{eq:Lreslimit}, below which there is no resonance. In this case $\mathcal{L}_{\rm reslim} = 1.83 \times 10^{-4}$  and the figure shows that for $\mathcal{L} > \mathcal{L}_{\rm reslim}$ there are two resonances, with the one at lower $T$ being dominant. For $\mathcal{L} = \mathcal{L}_{\rm reslim}$ there is only one resonance, and none for $\mathcal{L} < \mathcal{L}_{\rm reslim}$.

Inverting Eq.~\eqref{eq:Lreslimit} one can obtain the upper limit on $m_s$ to have a resonance for a given value of $\mathcal{L}$ and momentum $\epsilon$, called $(m_s)_\textrm{PEAK}$ in Ref.~\cite{Abazajian:2001nj},
\begin{equation}
\label{eq:mreslim}
    m_{\mathrm{reslim}} = (m_s)_\textrm{PEAK}= 4.03\times10^5 ~\textrm{keV}~ \epsilon^{-\frac{1}{2}}\mathcal{L}^{\frac{3}{2}}\left(\frac{B}{10.88\times10^{-9}\textrm{ GeV}^{-4}}\right)^{-1}~.
\end{equation}
Assuming $\epsilon= 1$ ($\epsilon$ is of $\mathcal{O}(1)$ or smaller for resonant production) for each particular lepton number in Figs.~\ref{fig:allreslim2} to \ref{fig:appresliml5} resonant production occurs only for $m_s < m_{\textrm{reslim}}$. No resonant production can occur in the region to the right of the vertical line labeled ``no res. prod." shaded in dark gray in the figures.

\subsection{Combined resonant and non-resonant production}
\label{ssec:resvsnonres}

When both $V_T$ and $V_D$ terms contribute to the potential of Eq.~\eqref{eq:mattermixing}, a combination of resonant and non-resonant production could occur at different times. It is thus important to consider the possibility of non-resonant production happening either before or after resonant production. As we will show, non-resonant production cannot happen before the resonant production. Such a scenario would result in a non-negligible initial $f_{\nu_s}$, which we assume to be zero. On the other hand, non-resonant production after the resonant production can result in a second population of sterile neutrinos, hotter than those produced resonantly. As we will discuss, there exists a narrow range of masses below $m_\textrm{reslim}$ in which this is possible, consistent with the previous findings of e.g. Refs.~\cite{Kishimoto:2008ic} and \cite{Boyarsky:2008mt}. 

Sterile neutrinos are produced non-resonantly  if the lepton number stored in active neutrinos is small enough so that the $V_D$ potential is negligible with respect to $V_T$, i.e. $V_T \gg V_D$. This can happen when an earlier resonant production has already depleted the active neutrino lepton asymmetry by converting it into sterile neutrinos, so that $V_D$ becomes small enough with respect to $V_T$.  In this case, there could be 
non-resonant production  after resonant production, for lower temperatures. This regime can be approximately identified by requiring that the temperature of maximum non-resonant production $T_{\rm max}$ (see Sec.~\ref{ssec:boltzmann})  is smaller  than the temperature of resonant production $T_{\rm res}$ (see Eq.~\eqref{eq:tres}). 
The condition $T_{\rm max} < T_{\rm res}$ translates into  a lower limit on the sterile neutrino mass $m_s > m_{\rm non-res}$. Note that because there is still considerable non-resonant production at temperatures below $T_{\rm max}$, $m_s \gtrsim m_{\rm non-res}$ is not a strict limit to have significant non-resonant production, which could also happen for masses smaller than but still close to $m_{\rm non-res}$.

For the standard cosmology, using $T_{\rm max}^{\rm Std}$ of Eq.~\eqref{eq:Stdmax}, we obtain
\begin{equation}
\label{eq:Stdmresmax}
m_s \gtrsim m_\textrm{non-res}^{\rm Std}  = 3.59\times10^{4}~\textrm{keV}~\epsilon_\textrm{res}^{\frac{3}{2}}\mathcal{L}^{\frac{3}{2}}~.
\end{equation}
Analogously, for the ST1 cosmology and using $T_{\textrm{max}}^{\rm ST1}$ of  Eq.~\eqref{eq:st1tmax} one obtains
\begin{equation}
\label{eq:st1resmax}
    m_s \gtrsim  m_\textrm{non-res}^{\rm ST1} = 6.11\times10^{4}~\textrm{keV}~\epsilon_\textrm{res}^{\frac{3}{2}}\mathcal{L}^{\frac{3}{2}}~.
\end{equation}
For the K and ST2 cosmologies,  $m_{\rm non-res}^{\rm K}$ and $m_{\rm non-res}^{\rm ST2}$ are given in App.~\ref{ssecapp:resvnonres}. 

\begin{figure*}[tb]
\mycenter{
\includegraphics[trim={0mm 0mm 0 0},clip,width=0.450\textwidth]{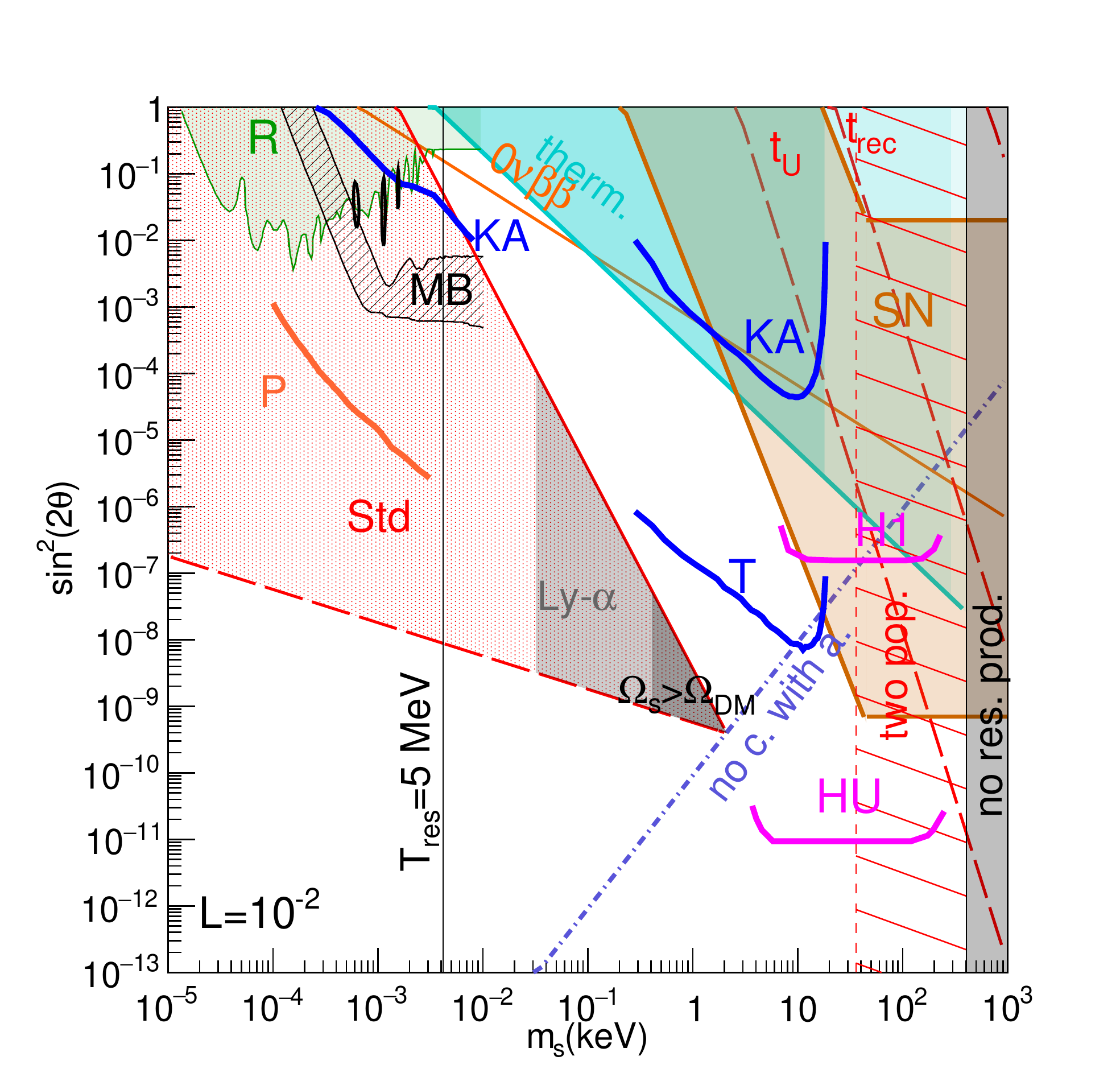}
\includegraphics[trim={0mm 0mm 0 0},clip,width=0.450\textwidth]{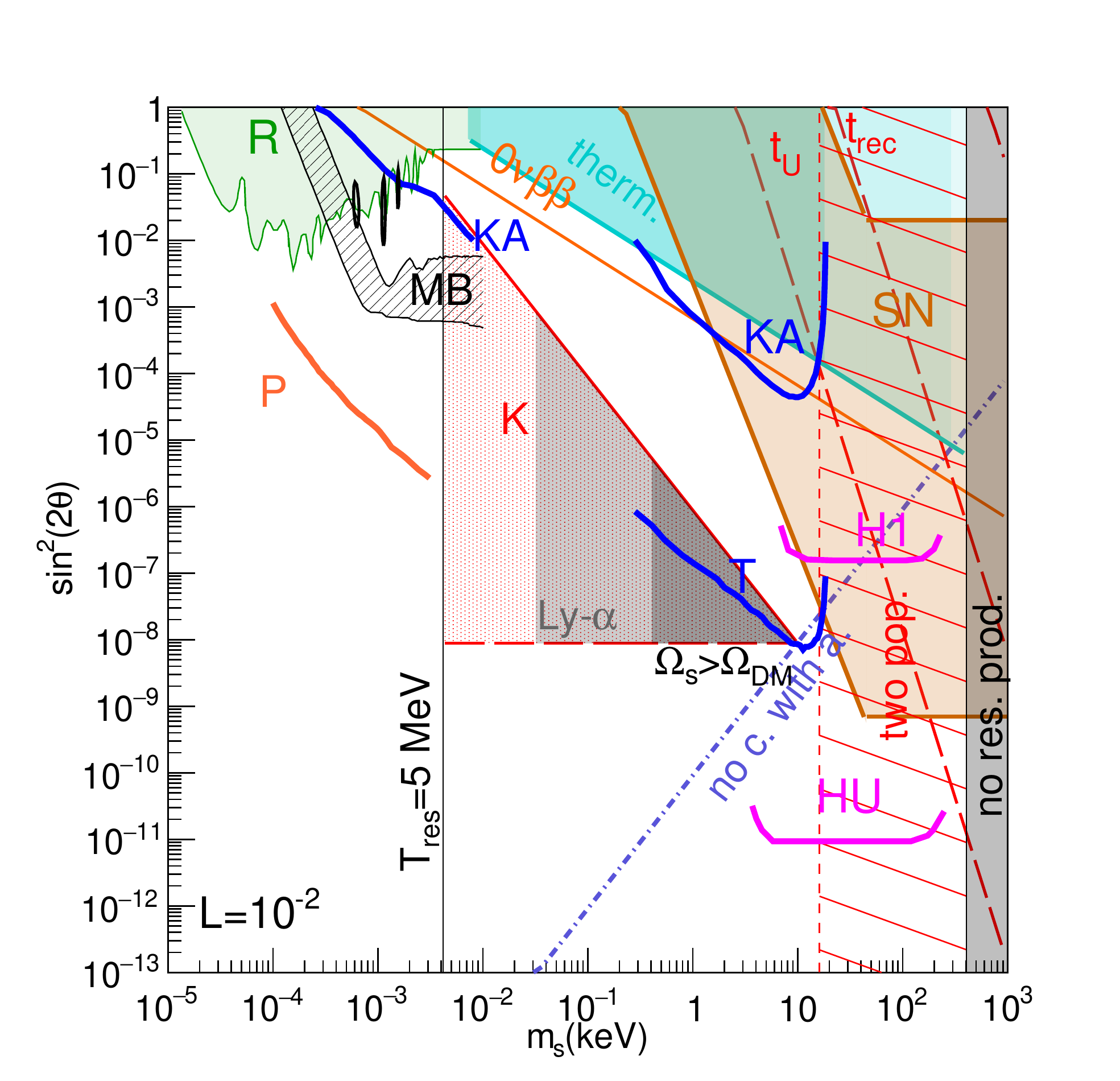}
}

\vspace{-0.2cm}

\mycenter{
\includegraphics[trim={0mm 0mm 0 0},clip,width=0.450\textwidth]{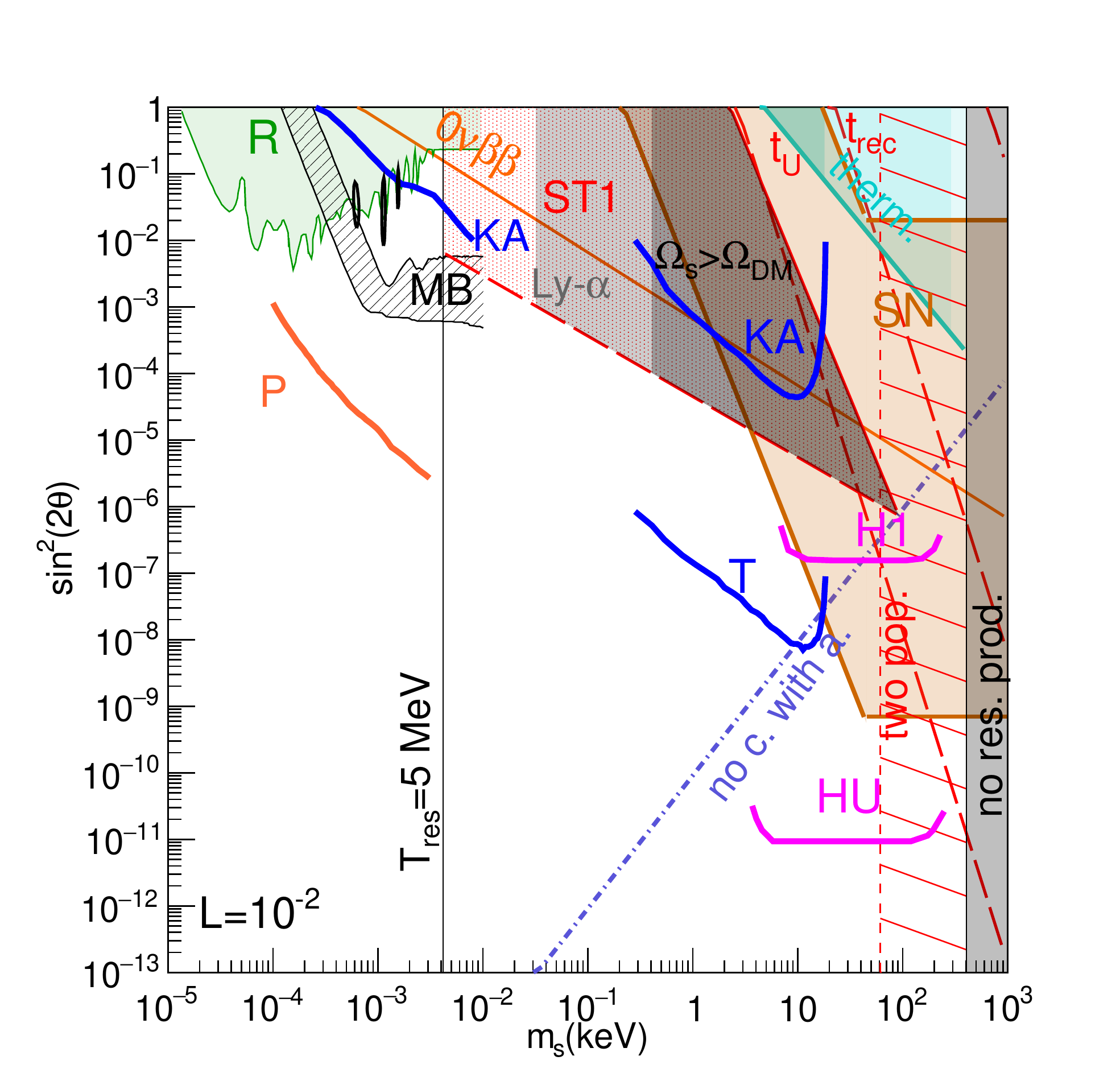}
\includegraphics[trim={0mm 0mm 0 0},clip,width=0.450\textwidth]{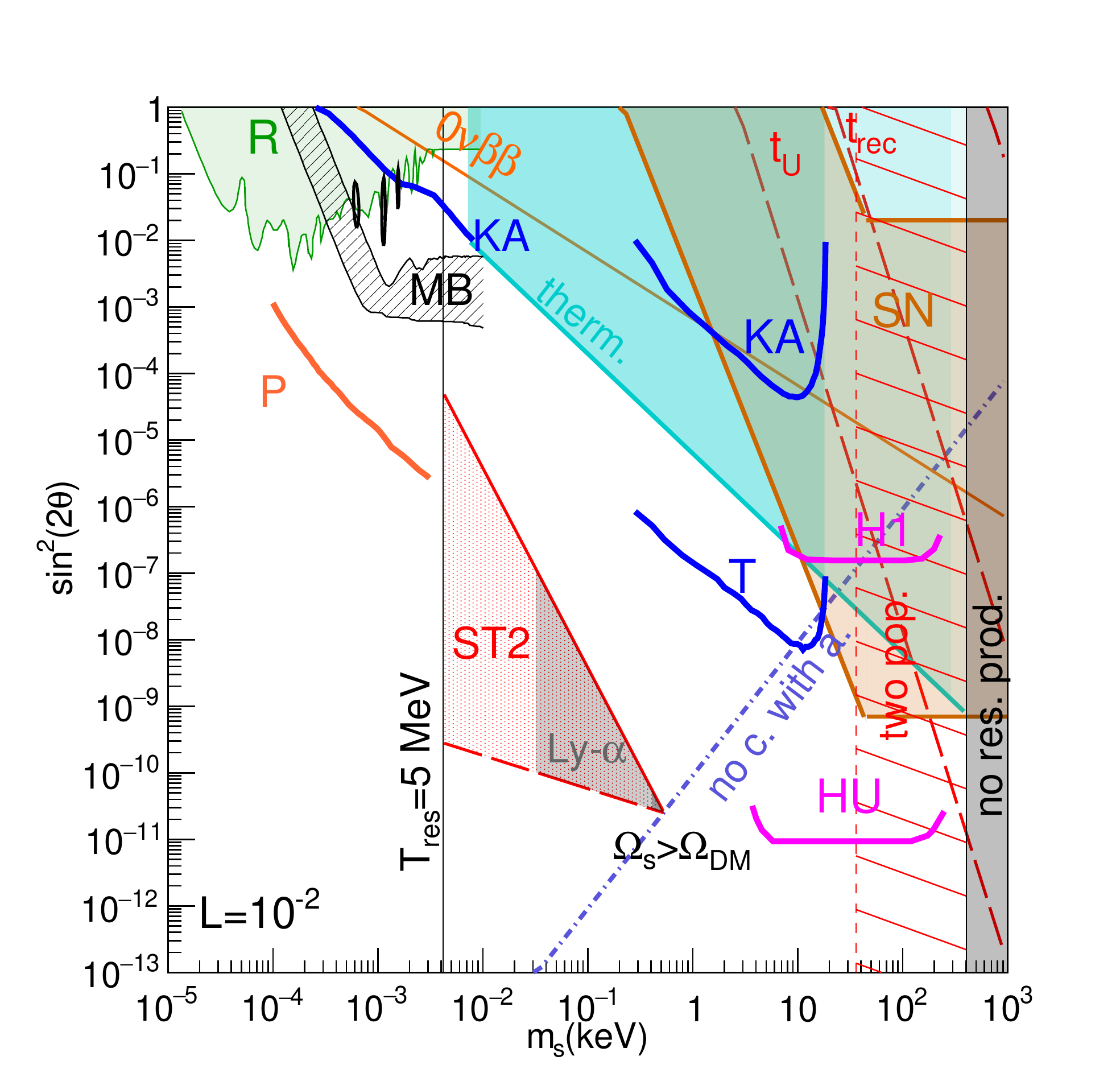}
}
\caption{\label{fig:allreslim2} 
{\small Regions/limits for resonantly produced $\nu_s$ mixed with $\nu_e$, for Std, K, ST1 and ST2 cosmologies assuming $\mathcal{L} = 10^{-2}$.  Resonant production is not possible to the right of the ``no res. prod.'' vertical black line (gray shading).  Additional non-resonant production is possible in the adjacent diagonally red hatched vertical strips. $T_{\rm res}<5$ MeV to the left of the  $T_{\rm res}$=5 MeV line.
Fully resonant conversion possible only above the violet dot-dashed diagonal line, within the red wedges between the adiabaticity and coherence limits. Regions excluded by $\Omega_s>\Omega_{\rm DM}$ and Ly-$\alpha$/HDM~\cite{Baur:2017stq} limits for fully resonantly produced $\nu_s$,  thermalization (``therm.") and reactor data (R)~\cite{An:2016luf, Declais:1994su, Ashenfelter:2018iov} are shaded in dark gray, gray, dark cyan and green respectively (lighter cyan ``therm." regions allowed due to entropy dilution). Shown are the upper limit from $0\nu\beta\beta$ decays (orange)~\cite{PhysRevLett.117.082503}, the  Majorana $\nu_s$ lifetimes $\tau_S= t_U$, $t_{\rm rec}$ and $t_{\rm th}$  (dashed red),  the reach of KATRIN (KA) and TRISTAN 3 yr (T)~\cite{Mertens:2018vuu, megas:thesis} in blue, HUNTER phase 1  (H1) and upgrade (HU)~\cite{Smith:2016vku} in purple, and PTOLEMY  for 100 g-yr (P) (Figs.~6 and 7 of~\cite{Betti:2019ouf}) in orange, the region (SN) disfavored by supernovae~\cite{Kainulainen:1990bn} (shaded in brown), the 4-$\sigma$ band of compatibility with LSND and MiniBooNE data (MB) in Fig.~4 of~\cite{Aguilar-Arevalo:2018gpe} (hatched in black)} and regions allowed at 3-$\sigma$ by DANSS~\cite{Alekseev:2018efk} and NEOS~\cite{Ko:2016owz} data in Fig.~4 of~\cite{Gariazzo:2018mwd} (3 black vertical contours).   
}
\end{figure*}

\begin{figure*}[h]
\mycenter{
\includegraphics[trim={0mm 0mm 0 0},clip,width=0.450\textwidth]{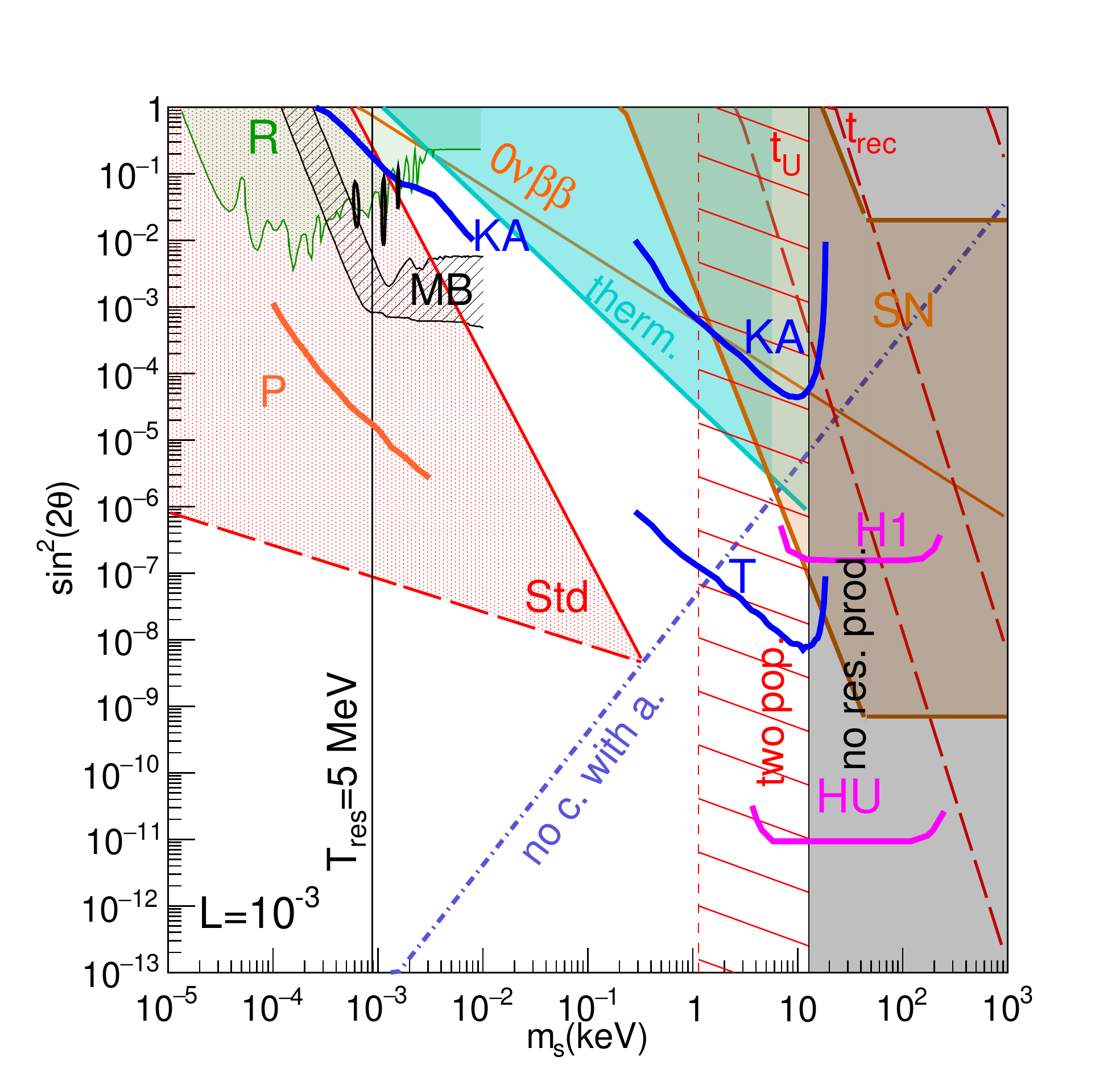}
\includegraphics[trim={0mm 0mm 0 0},clip,width=0.450\textwidth]{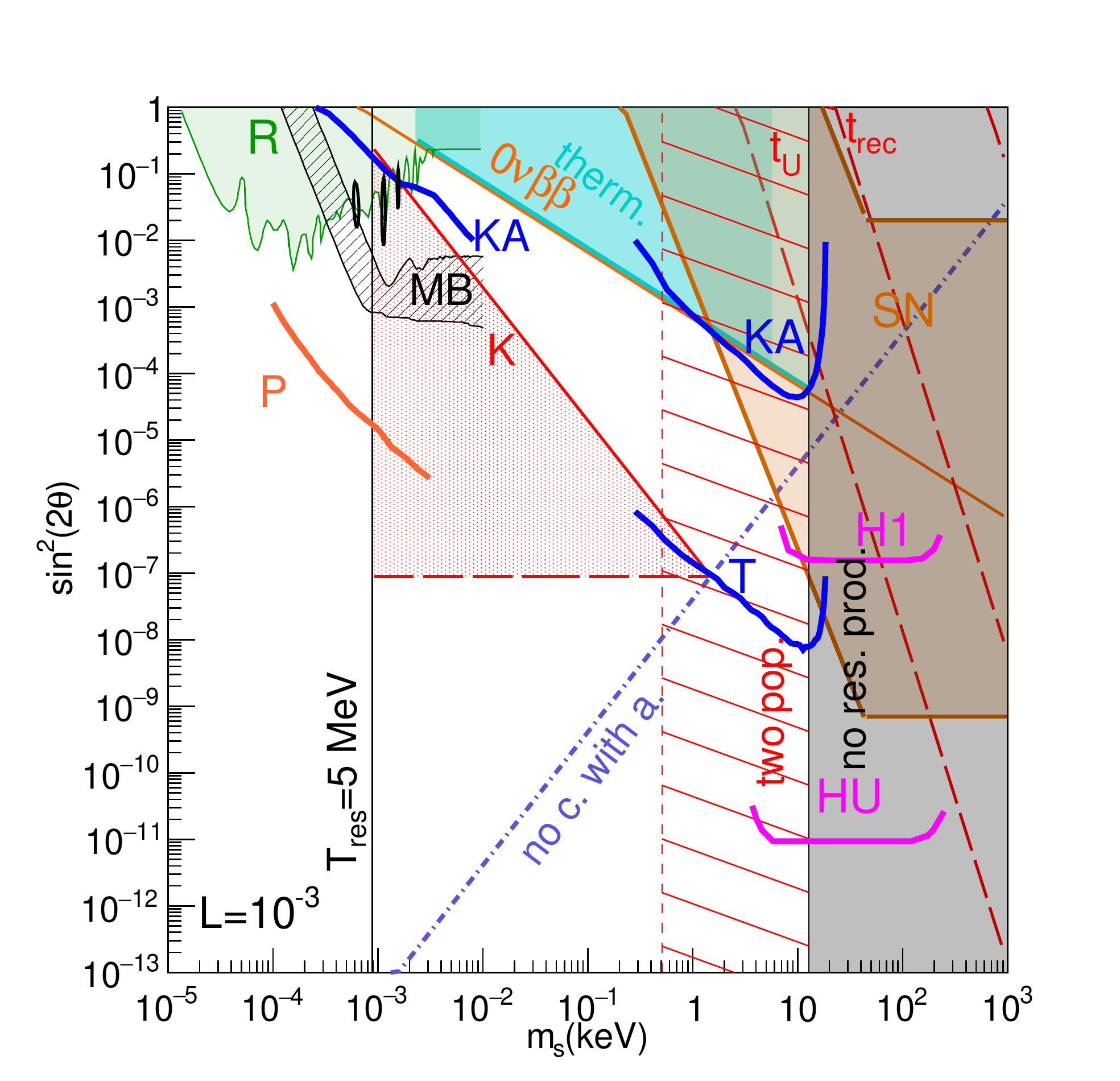}
}

\vspace{-0.2cm}

\mycenter{
\includegraphics[trim={0mm 0mm 0 0},clip,width=0.450\textwidth]{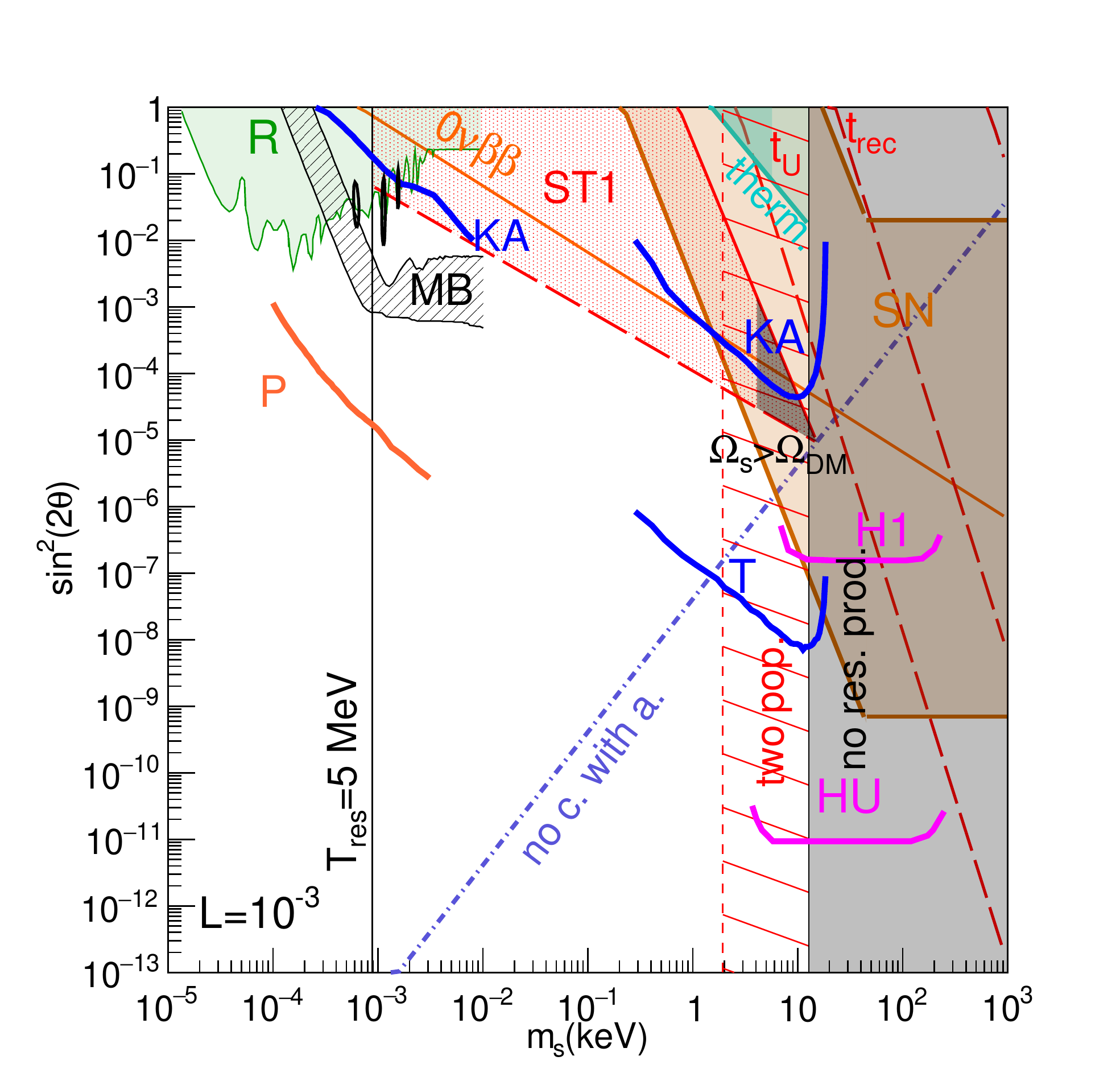}
\includegraphics[trim={0mm 0mm 0 0},clip,width=0.450\textwidth]{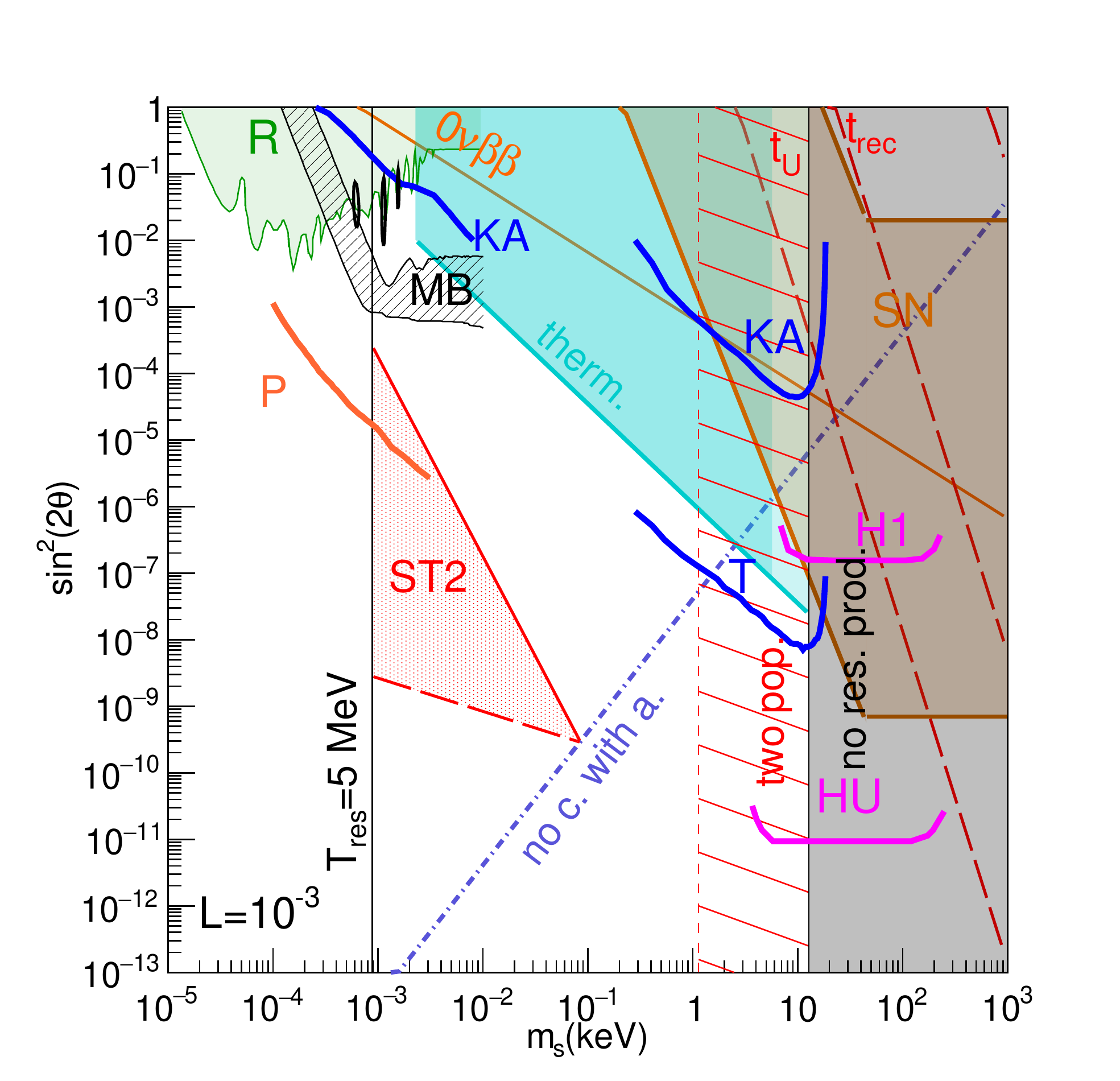}
}
\caption{\label{fig:appresliml3}
{\small As in Fig.~\ref{fig:allreslim2}, but for $\mathcal{L} = 10^{-3}$.}
}
\end{figure*}

\begin{figure*}[tb]
\mycenter{
\includegraphics[trim={0mm 0mm 0 0},clip,width=0.450\textwidth]{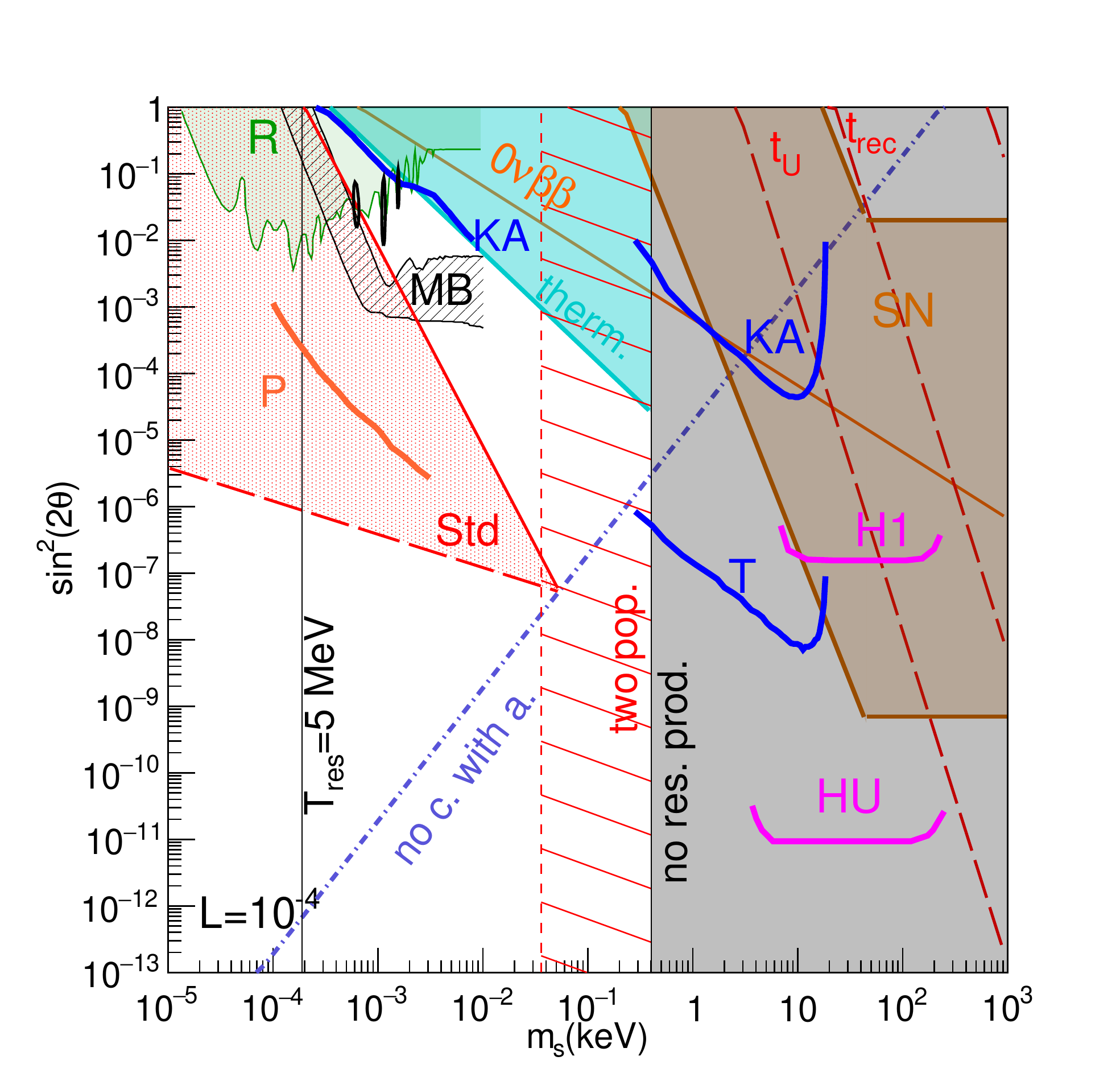}
\includegraphics[trim={0mm 0mm 0 0},clip,width=0.450\textwidth]{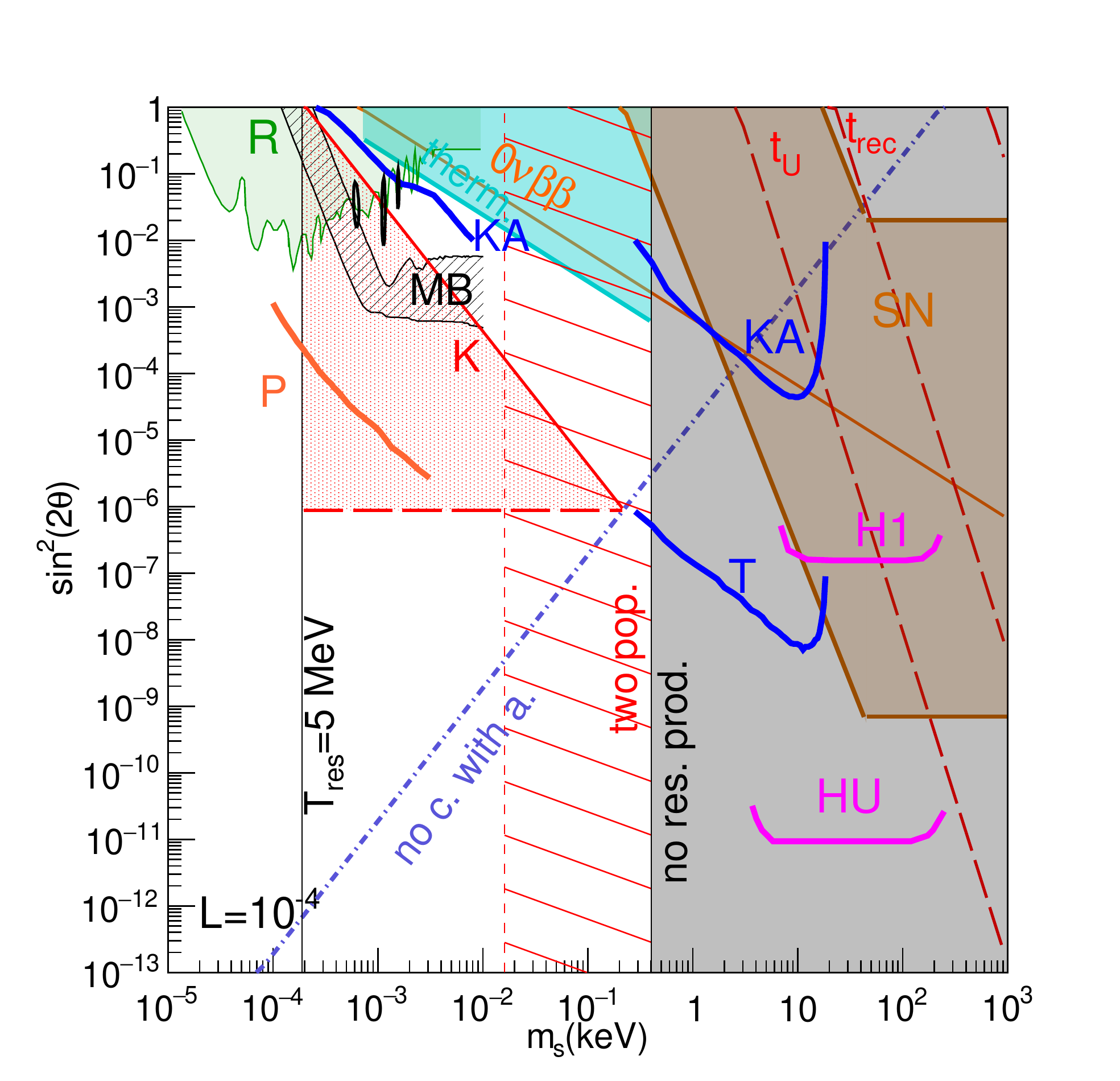}
}

\vspace{-0.2cm}

\mycenter{
\includegraphics[trim={0mm 0mm 0 0},clip,width=0.450\textwidth]{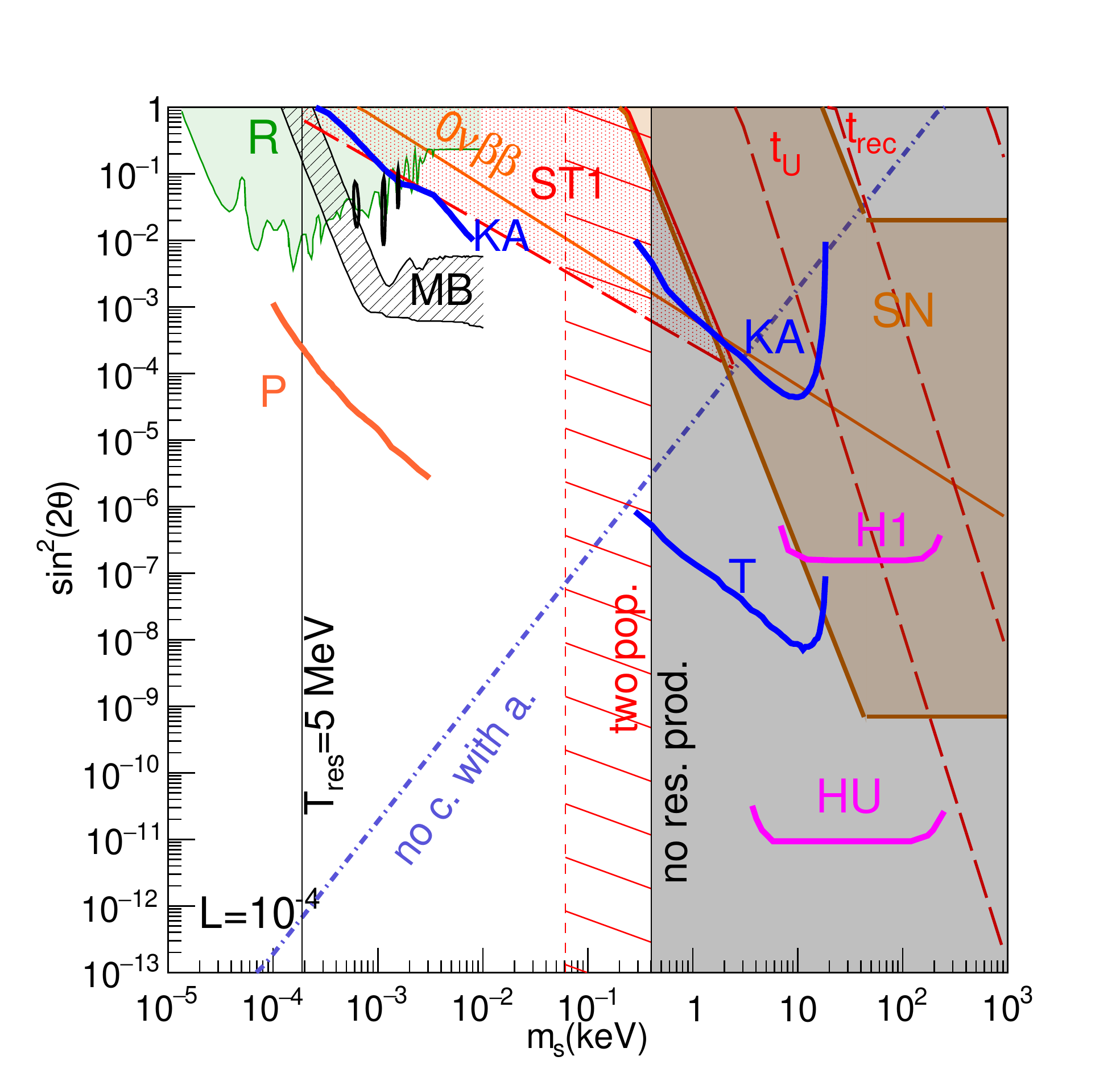}
\includegraphics[trim={0mm 0mm 0 0},clip,width=0.450\textwidth]{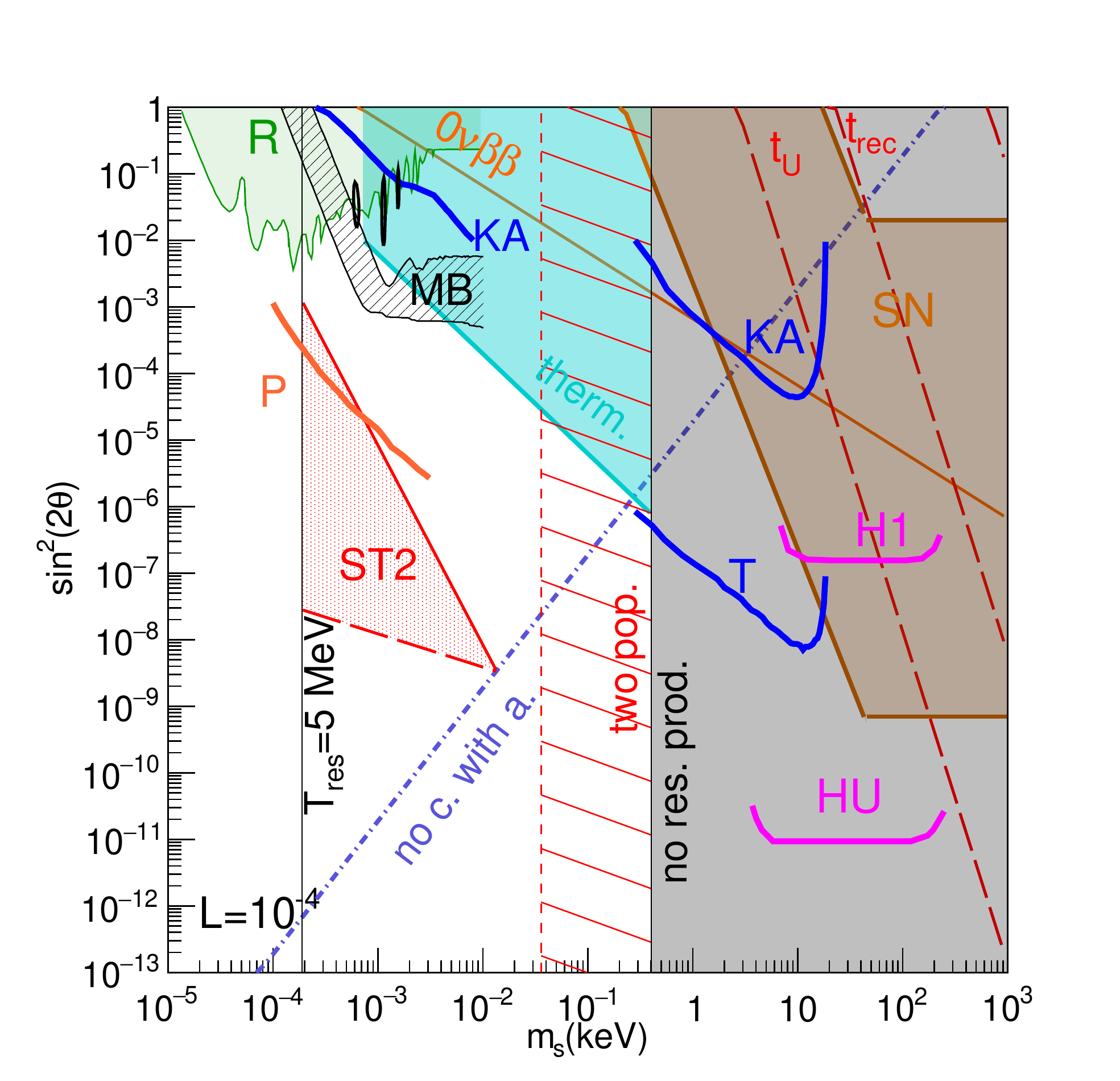}
}
\caption{\label{fig:allreslim1}
{\small As in Fig.~\ref{fig:allreslim2}, but for $\mathcal{L} = 10^{-4}$.}
}
\end{figure*}

\begin{figure*}[tb]
\mycenter{
\includegraphics[trim={0mm 0mm 0 0},clip,width=0.450\textwidth]{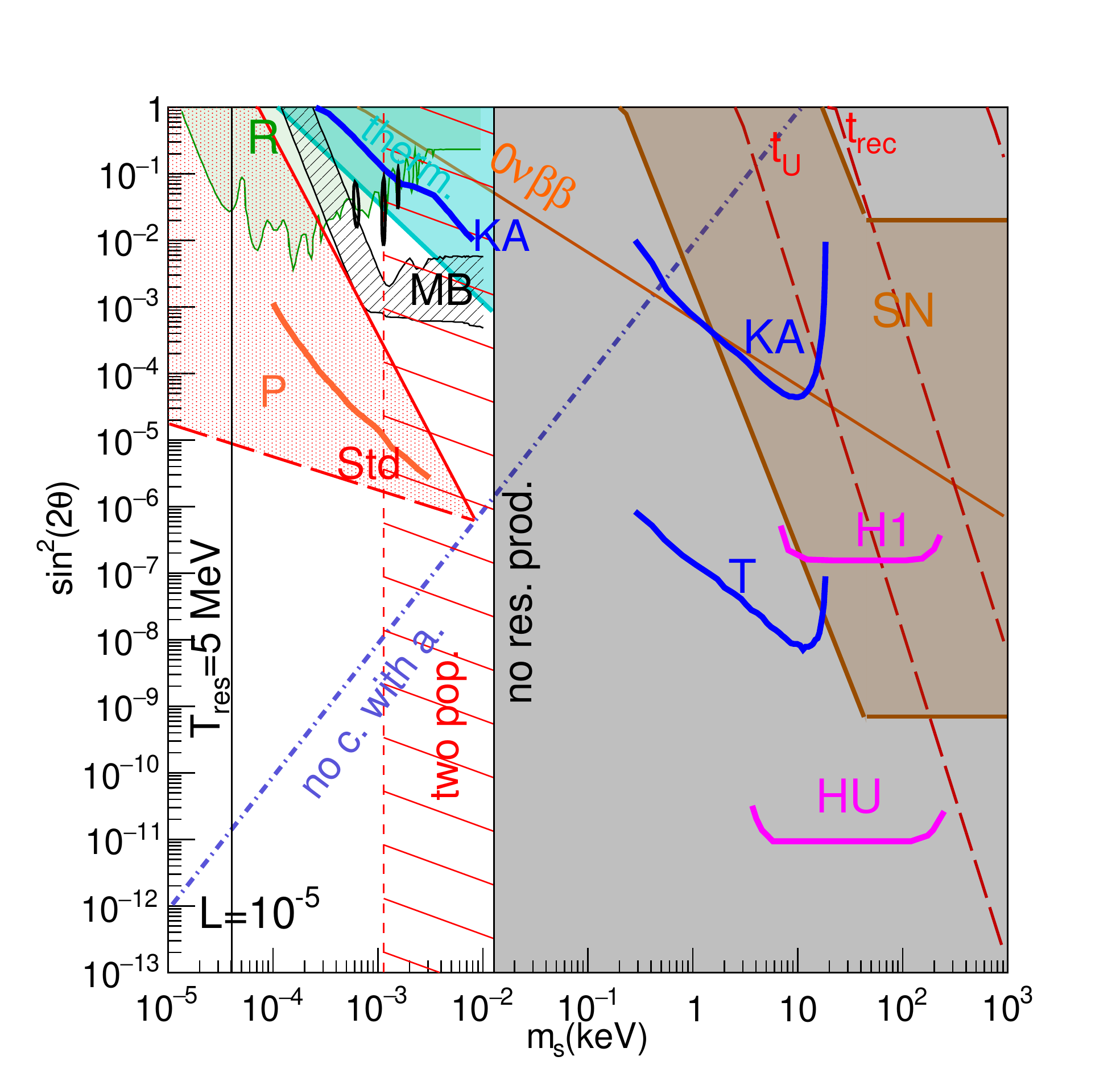}
\includegraphics[trim={0mm 0mm 0 0},clip,width=0.450\textwidth]{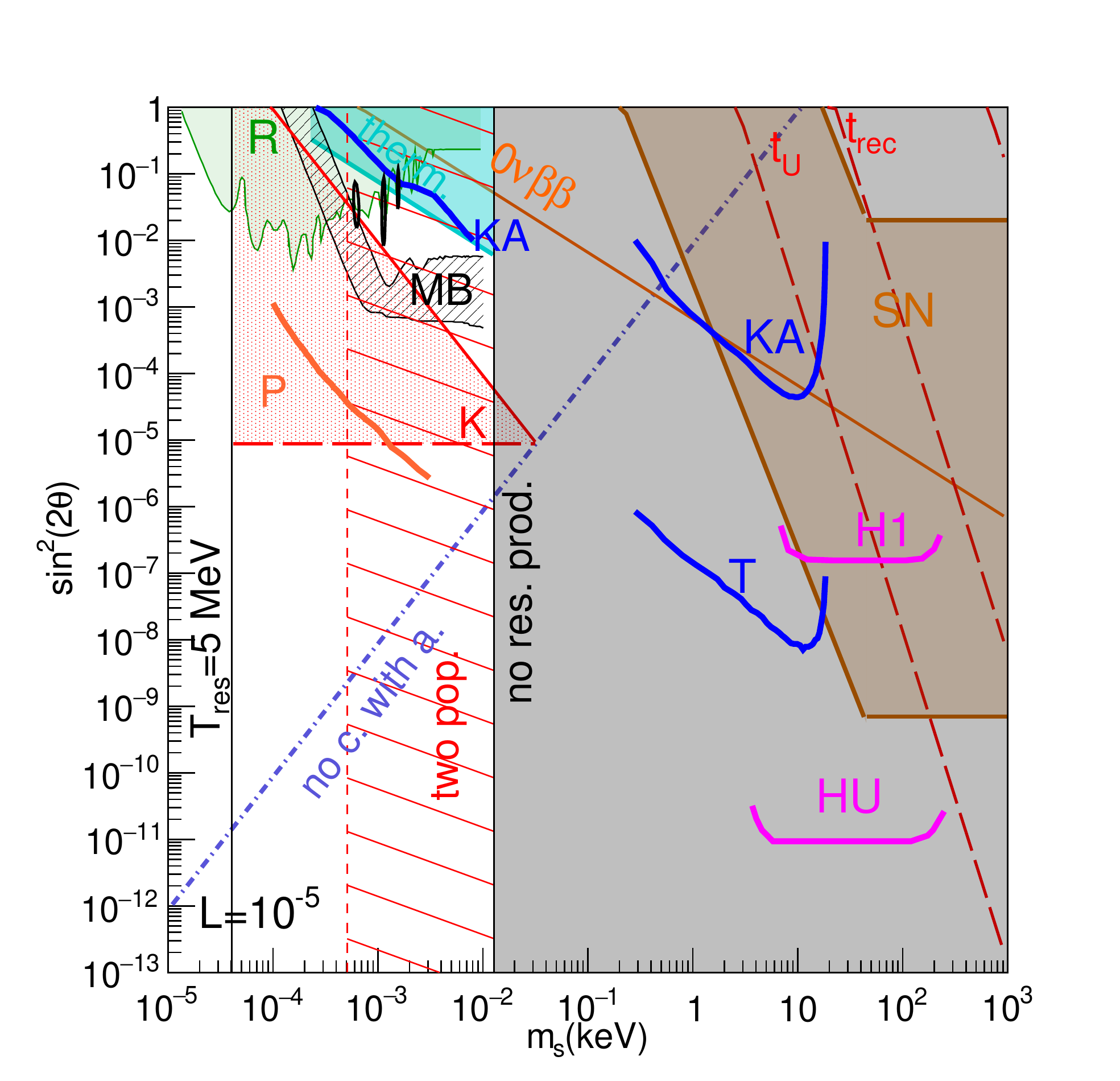}
}

\vspace{-0.2cm}

\mycenter{
\includegraphics[trim={0mm 0mm 0 0},clip,width=0.450\textwidth]{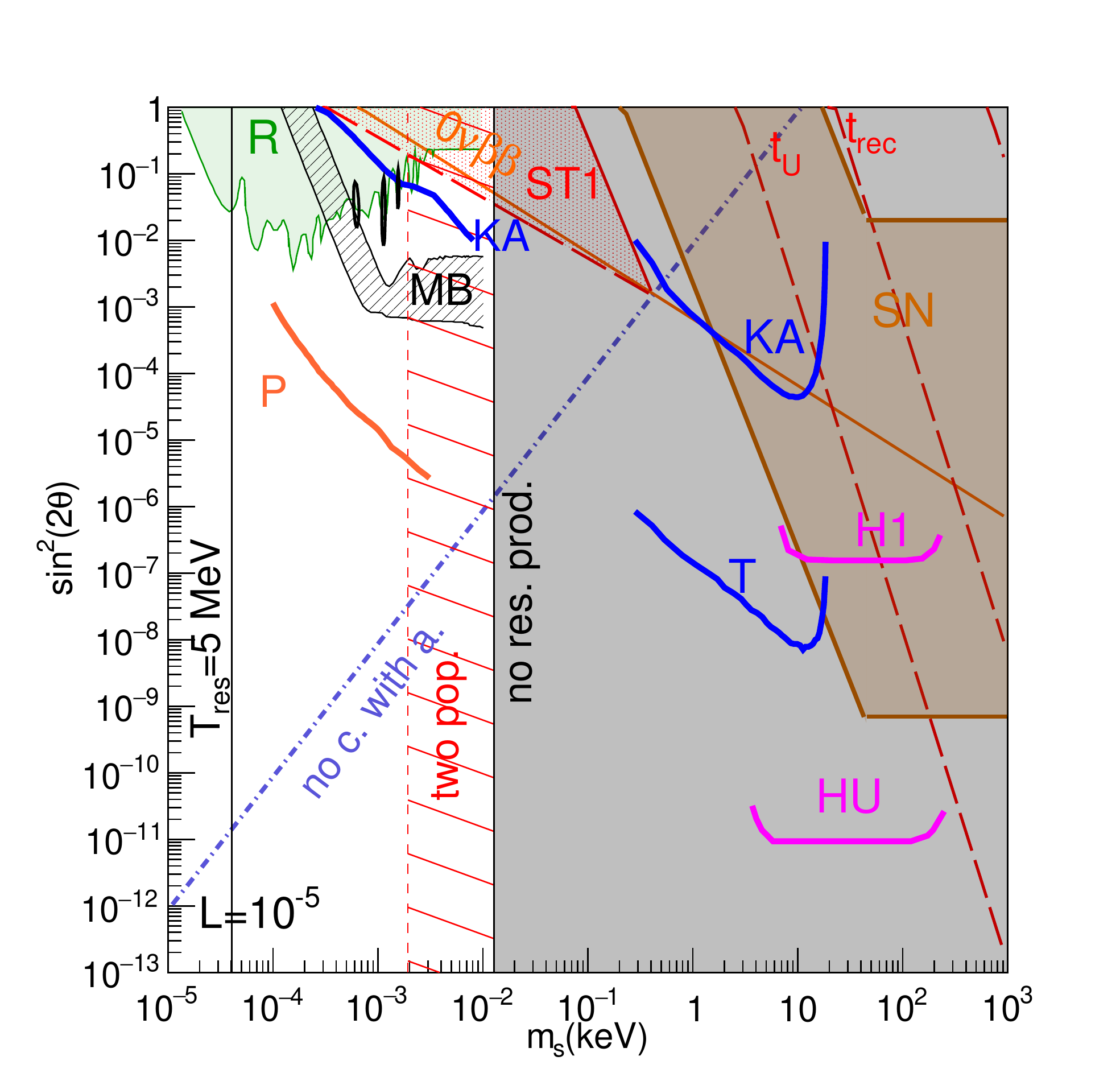}
\includegraphics[trim={0mm 0mm 0 0},clip,width=0.450\textwidth]{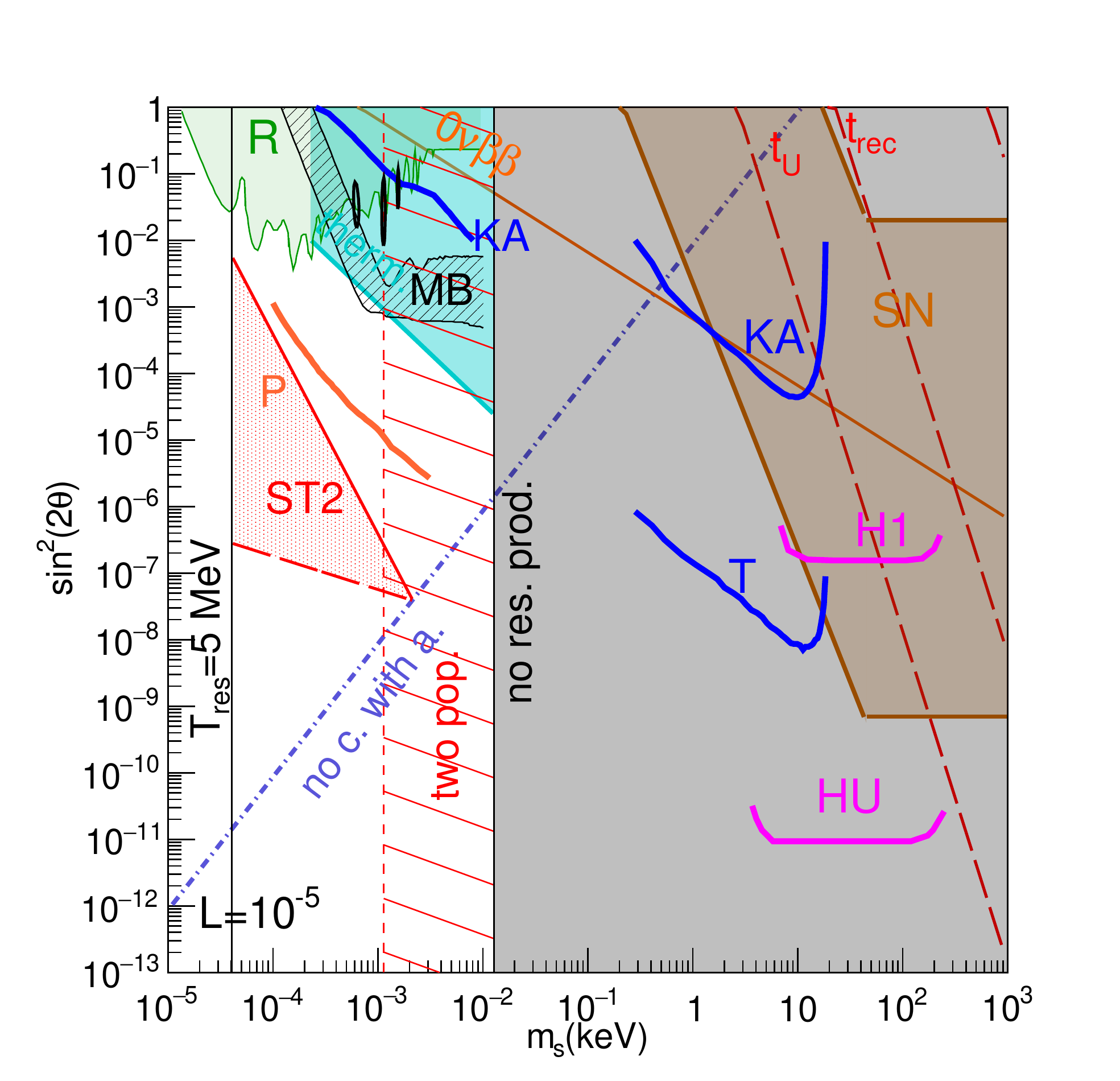}
}
\caption{\label{fig:appresliml5}
{\small As in Fig.~\ref{fig:allreslim2}, but for $\mathcal{L} = 10^{-5}$.}
}
\end{figure*}

As we have previously found, for a given lepton number $\mathcal{L}$ resonant production only occurs when  
$m_s < m_{\mathrm{reslim}}$  (see Eq.~\eqref{eq:mreslim}). Combining this upper limit with the lower limit above approximately identifies the range of  $m_s$ for each cosmology and each value of $\mathcal{L}$ where it is possible, following our arguments, to produce two distinct populations of the same sterile neutrino,
 \begin{equation} 
 \label{eq:msresrange}
     m_\textrm{non-res} \lesssim m_s<m_\textrm{reslim}~.
 \end{equation}
For $\epsilon =1$ this range defines the vertical bands diagonally hatched in red shown in Figs.~\ref{fig:allreslim2} to \ref{fig:appresliml5}, for  four different cosmologies and four values of the initial lepton number that we consider.  Note that full resonant production, where both the adiabaticity and coherence conditions discussed below are satisfied (shown as blue or red wedges in Figs.~\ref{fig:allreslim2} to \ref{fig:appresliml5}), seldom happens within the bands set by Eq.~\eqref{eq:msresrange}. 

Another condition for  two sterile neutrino populations to appear, assumed implicitly above, is that the initial lepton number
$\mathcal{L}$ in active neutrinos must be substantially depleted by the resonance. That is, the remaining lepton number after the resonant conversion, $\mathcal{L}_{\rm after}$,  must be small enough for the condition necessary for non-resonant production $V_T > V_D$ to hold. We will now show that this approximately requires $\mathcal{L}_{\rm after} \lesssim \mathcal{O}(0.1) \mathcal{L}$. 
 
 Consider that at  resonance the $V_D$ term (assuming $V_T$ is negligible) is equal to the mass term. On the other hand, for non-resonant production at $T_{\rm max}$ (Eq.~\eqref{eq:Tmaxstd} for the Std cosmology  and App.~\ref{ssecapp:maxtemp} for other cosmologies), the $V_T$ term is always approximately a fraction $\sim 0.34$ of the mass term\footnote{Here we used the definition of $T_{\rm max}$ which yields Eq.~\eqref{eq:Tmaxstd} (and the values in App.~\ref{ssecapp:maxtemp} for non Std cosmologies) and $\epsilon = 3.15$, which is the average value of $\epsilon$ resulting from DW production. We note that in Ref.~\cite{Gelmini:2019wfp,Gelmini:2019esj} a different  definition of $T_{\rm max}$ is used (not employing the momentum integrated rate and thus $\epsilon$ dependent), which results in the $V_T$ term being approximately 0.2  of the mass term at $T_{\rm max}$ instead. This does not significantly affect our arguments.}, assuming that the $V_D \ll V_T$. Since $V_D$  is linearly dependent on $\mathcal{L}$, immediately after the lepton number depletion at the resonance, the $V_D$ term diminishes by a factor of $(\mathcal{L}_{\rm after}/\mathcal{L})$. Thus, we need  $(\mathcal{L}_{\rm after}/\mathcal{L}) \lesssim 0.1$ in order to ensure $V_T > V_D$.
 This is often the case, e.g. the numerical simulations of resonant production of Ref.~\cite{Kishimoto:2006zk} find that $\gtrsim 90\%$ of the initial lepton number is depleted across a wide range of input parameters. The appearance of two sterile neutrino populations from non-resonant production that occurs after the resonant production, has been also found in some previous calculations, e.g. in Ref.~\cite{Kishimoto:2008ic}. For example, Fig.~1 of Ref.~\cite{Kishimoto:2008ic} shows that about $30\%$ of the sterile neutrino relic density is produced non-resonantly after significant lepton number depletion by resonant production. Using the parameters of that figure ($L_{\nu_e} = L_{\nu_\mu} = L_{\nu_\tau}=1.1 \times 10^{-3}$, i.e. $\mathcal{L}= 4.4 \times 10^{-3}$ and $m_s=64$ keV), we see that the assumed sterile neutrino mass is in between $ m_\textrm{non-res} \simeq 10$ keV and $m_\textrm{reslim} \simeq 117$ keV, for $\epsilon =1$ (a value of $\epsilon$ consistent with the resonant production shown in the same figure),  in agreement with our finding in Eq.~\eqref{eq:msresrange}. 
 
Two such populations of sterile neutrinos have been also found in Ref.~\cite{Boyarsky:2008mt} for $m_s \simeq $ O(keV) (e.g. for $m_s = 3$ keV in Fig.~1 of Ref.~\cite{Boyarsky:2008mt}), assuming the standard cosmology and a
``lepton asymmetry parameter''  $L_6 = 16$ (denoting $L= 16 \times 10^{-6}$).  As we explain now, these parameters appear to be consistent with Eq.~\eqref{eq:msresrange}, when expressed in terms of a range of $\mathcal{L}$ values for a fixed mass $m_s$. Using Eqs.~\eqref{eq:mreslim} and \eqref{eq:Stdmresmax} (taking the $B$-dependent factor to be 1) in Eq.~\eqref{eq:msresrange}, translates into the following condition for two populations to be produced,
\begin{equation} \label{eq:elrange}
    1.8\times 10^{-4} ~\epsilon^{1/2}~ \Big( \dfrac{m_s}{\text{keV}}\Big)^{2/3} < ~\mathcal{L}~ < 9.2 \times 10^{-4} ~ \epsilon^{-1}~\Big( \dfrac{m_s}{\text{keV}}\Big)^{2/3}~.
\end{equation}
 For $\epsilon \simeq 1$ and $m_s = 3$ keV (in agreement with what is shown in Fig.~1 of Ref.~\cite{Boyarsky:2008mt}), we see that $\mathcal{L}$ should be of order $10^{-4}$ for our condition of Eq.~\eqref{eq:elrange} to be fulfilled. It may seem contradictory then, that the ``lepton asymmetry parameter'' is $L_6 = 16$ in Fig.~1 of Ref.~\cite{Boyarsky:2008mt}. However, this parameter is defined as $L_6 = 10^6 (n_{\nu_e} - \bar{n}_{\nu_e})/s$ and, in the model used in Ref.~\cite{Boyarsky:2008mt} (see e.g. Ref.~\cite{Laine:2008pg}), there are 9 Weyl leptonic spinors with equal asymmetry $L_6$, which for us translates into $\mathcal{L} \simeq 9 (s/n_{\gamma})10^{-6} L_6 \simeq 1.44 \times 10^{-4} (s/n_{\gamma})$ for $L_6 = 16$. Additionally, the entropy density $s$ is larger than $n_{\gamma}$. Although a strict comparison of the results of Ref.~\cite{Boyarsky:2008mt} with ours is difficult to make due to the complexities of their model, the  lepton asymmetry specified in Ref.~\cite{Boyarsky:2008mt} appears to be compatible with our condition of Eq.~\eqref{eq:elrange} to have two sterile neutrino populations, both produced via active sterile neutrino oscillations.

 Let us now demonstrate that non-resonant sterile production does not occur prior to a resonant production. Resonant production after non-resonant production, i.e. $T_{\rm res} < T_{\rm max}$, would require that at some temperature the $V_D$ term is equal to the mass term (the condition for resonance) while at higher temperatures,  when both the $V_D$ and $V_T$ terms are larger (as both grow with increasing temperature) the  $V_T$  would be only $\sim 0.34~\times$ (mass term) and $V_D <V_T$ (the condition at maximum-non resonant production). This is not possible, since the mass term does not change with temperature. Hence, whenever $T> T_{\rm res}$ the mass term is guaranteed to be smaller than the $V_D$ term.
 
As an example of the preceding argument, let us consider the case of the standard cosmology. Using  $T_{\rm max}^{\rm Std}$ from Eq.~\eqref{eq:Stdmax}, the requirement of $V_D(T_{\rm max}) < V_T(T_{\rm max})$ translates into the condition 
\begin{equation} 
\label{eq:nonresmin}
    m_s >  5.5 \times 10^6~\text{keV} \epsilon^{-3/2} \mathcal{L}^{3/2}\left(\frac{B}{10.88\times10^{-9}\textrm{ GeV}^{-4}}\right)^{-1}~.
\end{equation}
However, this condition is incompatible  with having a resonance, $m_s < m_{\rm reslim}$ in  Eq.~\eqref{eq:mreslim}.  This implies
that significant non-resonant sterile neutrino production does not occur before resonant production, which is also true for the non-standard cosmologies we consider.

The considerations we discussed in this section thus far apply to all cases of resonant production in general. In the next subsection we focus specifically on fully resonant conversion.


\subsection{Fully resonant conversion}
\label{ssec:coherenceadiabaticity}

A ``fully resonant" neutrino production requires that two additional conditions hold during resonance: coherence and adiabaticity~\cite{Abazajian:2001nj}. If both are satisfied, there is a Mikheyev-Smirnov-Wolfenstein (MSW) resonance~\cite{Mikheev:1986gs, Wolfenstein:1977ue}. 

The first condition, coherence, ensures that neutrinos  do not suffer scattering-induced de-coherence  as they propagate  through the resonance. It is satisfied if the active neutrino mean free path (the inverse of the scattering rate $\Gamma_{\alpha}$ given above Eq.~\eqref{eq:interaction}) is significantly larger than the resonance width $\delta r_{\rm res}$, 
\begin{equation}
\label{eq:lres}
\Gamma_{\alpha}^{-1} >\delta r_{\rm res} \simeq \dfrac{ \tan(2 \theta)}{3 H}   \simeq \dfrac{\sin(2 \theta)}{3 H}~,
\end{equation}
where we took $\cos(2 \theta) \simeq 1$. This condition translates into an upper limit on the active-sterile mixing during resonance
\begin{equation}
 \label{eq:coherence}
    \sin^2 (2\theta) < \frac{9H^2}{G_{F}^{4} \epsilon^2 T_\textrm{res}^{10}}~,
\end{equation}
which defines the regions below the solid red lines in Figs.~\ref{fig:allreslim2} to \ref{fig:appresliml5} (below the upper boundary of the red shaded wedges).  Substituting  $T_{\rm res}$ from Eq.~\eqref{eq:tres} into Eq.~\eqref{eq:coherence} explicitly shows the dependence on the lepton number. The resulting expression for this upper limit written in terms of the $\eta$ and $\beta$ parameters  is given in Eq.~\eqref{eq:gencoherence}. In the Std cosmology this is 
\begin{equation}
\label{eq:Stdcoherence}
    \sin^2 (2\theta)^{\rm Std} <   2.19\times10^{-6}\epsilon^{-\frac{1}{2}}\left(\frac{m_s}{\textrm{keV}}\right)^{-3}\mathcal{L}^{\frac{3}{2}}\left(\frac{g_{\ast}}{10.75}\right)~,
\end{equation}
while for ST1 the upper limit is much higher 
\begin{equation}
\label{eq:st1coherence}
   \sin^2 (2\theta)^{\rm ST1} < 1.38\times10^{5} \epsilon^{-0.09}\left(\frac{m_s}{\textrm{keV}}\right)^{-3.82}\mathcal{L}^{1.91}\left(\frac{T_\textrm{tr}}{5\textrm{ MeV}}\right)^{1.64}\left(\frac{g_{\ast}}{10.75}\right) ~.
\end{equation}
Results for K and ST2 are given in Appendix~\ref{ssecapp:coherence}.

When the neutrino flavor evolution is adiabatic, neutrinos oscillate many times during the resonance width, in which case flavor conversion is very efficient and the lepton number in active neutrinos is quickly transferred to sterile neutrinos. Comparing the neutrino oscillation length $l_m^{\rm res}$ in matter at the resonance
\begin{equation}
l_m^{\rm res} = \dfrac{4 \pi \epsilon T_{\rm res} (\epsilon)}{m_s^2 \sin(2 \theta)}
\end{equation}
with the resonance width $\delta r_{\rm res}$ from Eq.~\eqref{eq:lres}, the degree of adiabaticity can be characterized using the $\gamma$ parameter (see e.g. \cite{Abazajian:2001nj}),
\begin{equation}
\label{eq:adiab}
\gamma = 2 \pi \frac{\delta r^{\rm res}}{l_m^{\rm res}} \simeq \frac{m_s^2}{2\epsilon T_\textrm{res}} \sin^2 (2\theta)\left|4\frac{\dot{T}}{T}+\frac{\dot{\mathcal{L}}}{\mathcal{L}}\right|^{-1} =~ \frac{m_s^2}{2\epsilon T_\textrm{res}} \sin^2 (2\theta)\left|4 H-\frac{\dot{\mathcal{L}}}{\mathcal{L}}\right|^{-1}~.
\end{equation}
The condition $\gamma > 1$ for adiabatic evolution ensures that the change in mass eigenstates proceeds slowly  during level crossing at the resonance, so that the probability of jumping between mass eigenstates is low\footnote{Assuming a linear change in potential across resonance and small mixing angles, 
the probability of $\nu_{\alpha} \rightarrow \nu_s$ conversion at resonance is $P_{\nu_s \rightarrow \nu_\alpha} = 1-P_\textrm{LZ}$, where
$P_\textrm{LZ} \simeq \exp(-\pi\gamma /2)$  is the Landau-Zener probability~\cite{Abazajian:2001nj}.}. Eq.~\eqref{eq:adiab} shows that for an evolution initially described by $\gamma \gg 1$  to not lead to one with $\gamma < 1$, the sweep  $(\dot{\mathcal{L}}/\mathcal{L})$ should remain small, not much larger than $H$ (see the discussion in Ref.~\cite{Kishimoto:2006zk}).

If $(\dot{\mathcal{L}}/\mathcal{L}) < H $, we can neglect $(\dot{\mathcal{L}}/\mathcal{L})$ in Eq.~\eqref{eq:adiab}, and the adiabaticity condition $\gamma > 1$ reduces to
\begin{equation}
\label{eq:adiabaticity2}
    \sin^2(2\theta) > \frac{8\epsilon H T_\textrm{res}}{m_s^2}~.
\end{equation}
This condition defines the mass-mixing regions above the long dashed red straight lines in Figs.~\ref{fig:allreslim2} to \ref{fig:appresliml5} (above the lower boundary of the red shaded  wedges). In terms of general parameters $\eta$ and $\beta$,  Eq.~\eqref{eq:adiabaticity2} is given in Eq.~\eqref{eq:genadiabaticity}. In the Std cosmology, this condition is
\begin{equation}
\label{eq:Stdadiabaticity}
    \sin^2(2\theta)^{\rm Std} > 2.34\times10^{-11}\gamma_\textrm{lim}\epsilon^\frac{1}{4}\left(\frac{m_s}{\textrm{keV}}\right)^{-\frac{1}{2}}\mathcal{L}^{-\frac{3}{4}}\left(\frac{g_{\ast}}{10.75}\right)^{\frac{1}{2}}~,
\end{equation}
where $\gamma_\textrm{lim}$ is the chosen minimum value of the parameter $\gamma$  (we take $\gamma_\textrm{lim}=1$ for our figures, but in general $\gamma_\textrm{lim}\geq 1$). For the ST1 cosmology, the condition becomes
\begin{equation}
\label{eq:st1adiabaticity}
    \sin^2(2\theta)^{\rm ST1}>5.88\times10^{-6}\gamma_\textrm{lim}\epsilon^{0.46}\left(\frac{m_s}{\textrm{keV}}\right)^{-0.91}\mathcal{L}^{-0.55}\left(\frac{g_{\ast}}{10.75}\right)^{\frac{1}{2}}\left(\frac{T_\textrm{tr}}{5\textrm{ MeV}}\right)^{0.82}~.
\end{equation}
Expressions for the K and ST2 cosmologies are provided in the Appendix~\ref{ssecapp:adiabaticity}.
 
As the resonance sweeps across the active neutrino momentum distribution $\mathcal{L}$ decreases, while the fractional change in lepton number $(\dot{\mathcal{L}}/\mathcal{L})$  steadily increases. When $(\dot{\mathcal{L}}/\mathcal{L}) \gg H$, as seen from Eq.~\eqref{eq:adiab}, $\gamma$ rapidly decreases and adiabaticity is lost. 

The maximum value of $\epsilon$, $\epsilon_{\rm max}$, can be found from the condition~\cite{Kishimoto:2006zk}
\begin{equation}
\label{eq:emax}
    \mathcal{L}(\epsilon_{\rm max}) ~=~ \frac{1}{2\zeta(3)}\left(\frac{T_{\nu}}{T_\gamma}\right)^3\frac{\epsilon_\textrm{max}^3}{(e^{\epsilon_\textrm{max}-\xi}+1)}
    ~=~ \mathcal{L}_\textrm{init} - \frac{1}{2\zeta(3)}\left(\frac{T_{\nu}}{T_\gamma}\right)^3 \int_0^{\epsilon_\textrm{max}}\frac{x^3}{e^{x-\xi}+1}dx~.
\end{equation}
To solve this equation we neglect $\xi$, having verified that $\xi \ll \epsilon_{\rm max}$. The momentum distribution of resonantly produced sterile neutrinos has a sharp peak at $\epsilon \simeq \epsilon_{\rm max}$ (see e.g. Fig.~2 of Ref.~\cite{Kishimoto:2006zk}). Hence, we can estimate that the average $T$-scaled momentum is  approximately\footnote{We note that this approximation may not entirely capture the relevant behavior if the distribution has a long tail in one direction.} $\langle \epsilon \rangle \simeq \epsilon_{\rm max}$.

To obtain the region in sterile neutrino mass and mixing parameter space associated with adiabaticity and coherence and to compute the limits in Sec.~\ref{sec:limits} that depend on the relic abundance, we numerically solve  Eq.~\eqref{eq:emax} for $\epsilon_{\rm max}$ and then set $\epsilon = \epsilon_{\rm max}$ in Eq.~\eqref{eq:coherence} and Eq.~\eqref{eq:adiabaticity2}. Eq.~\eqref{eq:emax} shows that $\epsilon_{\rm max}$ depends only on the initial value of $\mathcal{L}$, thus is the same for all the cosmologies we consider. For the values of $\mathcal{L}$ used in Figs. \ref{fig:allreslim2} to  \ref{fig:appresliml5}, we find that $\epsilon_{\rm max}$ is
\begin{equation}
\label{eq:reseps}
\langle \epsilon\rangle \simeq \epsilon_{\rm max}
= \left \{
  \begin{tabular}{cl}
  0.03, & ~~~${\rm for}~~ \mathcal{L}=10^{-5}$\\
  0.07, & ~~~${\rm for}~~\mathcal{L}=10^{-4}$ \\
  0.35, & ~~~${\rm for}~~\mathcal{L}=10^{-3}$ \\
  0.83, & ~~~${\rm for}~~\mathcal{L}=10^{-2}$~. \\
  \end{tabular}
\right .\
\end{equation}

For each particular cosmology we consider,  
the coherence and adiabaticity conditions determine a wedge shown shaded in red for the Std, ST1, K and ST2 cosmologies in the mass-mixing plane  in Figs.~\ref{fig:allreslim2} to \ref{fig:appresliml5}, between the solid and dashed red boundary straight lines, where fully resonant MSW conversion happens. The particular shape of the wedge delineating the full resonance conversion region is governed by the specific temperature dependence of $H(T)$ in each cosmology (i.e. the $\beta$ dependence in Eq.~\eqref{eq:hnStd}), while the location of the wedge depends on the $\eta$ parameter. Since the coherence condition (upper solid boundary line) of Eq.~\eqref{eq:coherence} scales as $H^2$ and the adiabaticity condition (lower dashed boundary line) of Eq.~\eqref{eq:adiabaticity2} scales as $H$,  the wedges of non-standard cosmologies with higher expansion rates compared to Std move to larger mixings, and those with lower expansion rates move to smaller mixings.  

Equating the conditions for coherence and adiabaticity,  Eq.~\eqref{eq:lres} and Eq.~\eqref{eq:adiab} with $\gamma=1$, gives
\begin{equation}
\label{eq:intersect}
  \Gamma_\alpha  l_m^{\rm res} = 2 \pi~. 
\end{equation}
This corresponds to the locus of intersection of the adiabaticity  $\gamma = 1$ boundary (lower  dashed boundary of the wedges in the figures) and the coherence $\Gamma_{\alpha}^{-1} = \delta r_{\rm res}$  boundary (the upper solid boundary of the wedges). The line of intersections is valid for all cosmological models and is defined by the following mixing angle as a function of mass 
\begin{equation}
\label{eq:intersections}
    \sin^2 (2\theta)_{\rm intersect}  = 2.60\times10^{-16}\epsilon   \left(\frac{m_s}{\textrm{keV}}\right)^2   \mathcal{L}^{-3}~,
\end{equation}
which is indicated with a diagonal dot-dashed violet line in Figs.~\ref{fig:allreslim2} and \ref{fig:allreslim1}. Hence, coherence (i.e. $\Gamma_{\alpha}^{-1} >\delta r_{\rm res}$) and adiabaticity (i.e. $\delta r_{\rm res} > {l_m^{\rm res}}/{2 \pi}$)  can only happen together for $\sin^2 (2\theta) > \sin^2 (2\theta)_{\rm intersect}$, where $\Gamma_{\alpha}^{-1} > {l_m^{\rm res}}/{2 \pi}$.

As we mentioned above, in the right hand side of Eq.~\eqref{eq:boltzmann2} we omitted the quantum damping factor $[1- (\Gamma_\alpha \ell_m/2)^2]^{-1}$  since it is generally negligible for the range of parameters relevant for our study. However, for $\sin^2 (2\theta) < \sin^2 (2\theta)_{\rm intersect}$, this factor becomes non-negligible at the resonance, where the neutrino oscillation length $\ell_m^{\rm res}$ is maximal. Namely, we have $\ell_m^{\rm res} > 2 \pi \Gamma_\alpha^{-1}$ in the parameter space below and to the right of the diagonal violet dot-dashed line labelled ``no c. with a." (i.e. ``no coherence with adiabaticity") in Figs.~\ref{fig:allreslim2} to \ref{fig:appresliml5}. In this region, neutrino scattering-induced decoherent production takes place (see discussion in Ref.~\cite{Kishimoto:2008ic}).

Assuming that fully resonant conversion has happened and that most of the initial lepton number 
$\mathcal{L}$ has been depleted,
the  present number density of resonantly produced sterile neutrinos, $n_{\nu_s, {\rm res}}$, is  given by
\begin{equation}
\label{eq:SFproduction}
n_{\nu_s, {\rm res}} = \Big(\sum L_{\nu_{\alpha}}\Big) n_{\gamma} = \left(\frac{3}{4}\mathcal{L}\right) \frac{2\zeta(3)}{\pi^2}T_0^3~.
\end{equation}
Here we have used Eq.~\eqref{eq:3/4L} to rewrite $\sum L_{\nu_{\alpha}}$ in terms of $\mathcal{L}$, assuming that neutrino oscillations redistribute efficiently the lepton asymmetry among the flavors, and $T_0 = 2.75$~K is the present temperature of the CMB. Using this equation, the fraction of the DM consisting of resonantly produced sterile neutrinos is 
\begin{equation}
\label{eq:sfoverdensity}
    f_{s, {\rm DM}}^{\rm res} = \frac{n_{\nu_s, {\rm res}}~ m_s} {\rho_{\rm DM}} = \left(\dfrac{m_s}{4.08~\text{eV}}\right)  \left(\frac{ 0.12}{\Omega_\textrm{DM}h^2}\right) \mathcal{L}~.
\end{equation}
Thus, requiring $f_{s, {\rm DM}}^{\rm res}\leq 1$ implies an upper limit on the sterile neutrino mass.
In Figs.~\ref{fig:allreslim2} to~\ref{fig:appresliml5}, the portions of the wedges where sterile neutrinos can be fully resonantly produced in which $f_{s, {\rm DM}}^{\rm res} > 1$ are
shaded in dark gray and labelled $\Omega_s>\Omega_{\rm DM}$. 

We stress two important assumptions that go into our estimation of $f_{s, {\rm DM}}^{\rm res}$, namely (1) that the conversion is adiabatic and coherent (which does not hold outside the wedges shaded in red, for each particular considered cosmology and lepton number, in Figs.~\ref{fig:allreslim2} to~\ref{fig:appresliml5}) and (2) that practically all of the initial lepton number is converted into sterile neutrinos. If only a small fraction of the initial lepton number is converted into sterile neutrinos, their resulting number and energy densities would be reduced by the same fraction. It is important to acknowledge these approximations when comparing our abundance limits to the existing literature that deals with resonant production of sterile neutrinos, e.g. Ref.~\cite{Abazajian:2014gza}.
For fully resonant conversion the number density is fixed solely by the lepton number and thus the relic energy density is independent of the mixing angle, in contrast to what is shown in Ref.~\cite{Abazajian:2014gza}. 

\subsection{Thermalization}
\label{ssec:therm}

Let us consider the possibility that production of sterile neutrinos due to interactions with the thermal bath
could have brought the sterile neutrinos to thermal equilibrium before resonant production could occur. In this case, without significant entropy dilution, sterile neutrinos could contribute to the energy density as much as an active neutrino species, i.e. $\Delta N_{\rm eff} = 1$, during BBN and thus be forbidden by  the limit $N_\textrm{eff}<3.4$~\cite{Tanabashi:2018oca} (see below). The thermal dominant production rate is $\Gamma \simeq \sin^2(\theta_m) \Gamma_\alpha \simeq \sin^2(\theta_m)d_\alpha G_F^2\epsilon T^5$ (similar to the rate in Eq.~\eqref{eq:interaction}), where we have used Eq.~\eqref{eq:interaction-active} for the active neutrino rate $\Gamma_\alpha$ and assumed that one sterile neutrino is produced in each interaction instead of an active one.
This rate can be suppressed by a large enough lepton number $\mathcal{L}$,  because the corresponding density potential $V_D$ diminishes the  mixing angle in the medium $\sin (\theta_m)$. This  could prevent thermalization even for relatively high vacuum mixing angles. We derive below a conservative upper bound, above which thermalization should occur, and compare it with Fig. 4 of Ref.~\cite{Hannestad:2012ky} where the same limit is computed numerically for $m_s \simeq eV$ solving quantum kinetic equations. The parameter space above this bound is forbidden, and possibly, some of the space below near it could be forbidden too by a more accurately derived limit.

 The denominator of the matter mixing angle  in Eq.~\eqref{eq:mattermixing} is dominated by the $V_D$ term   for temperatures in between the two resonance temperatures, namely   when $2\epsilon T V_D /m_s^2> 1$,  i.e. for temperatures above the lower-temperature resonance, and $2\epsilon T V_D /m_s^2 > 2\epsilon T V_T /m_s^2$   i.e. for temperatures below the higher-temperature resonance.   These two conditions give the range of temperatures where the density potential $V_D$ is dominant, 
\begin{equation}
\label{eq:thermtemprange}
   0.596\textrm{ MeV}\epsilon^{-\frac{1}{4}}\left(\frac{m_s}{\textrm{eV}}\right)^\frac{1}{2}\mathcal{L}^{-\frac{1}{4}} <  T < 19.1\textrm{ GeV}\epsilon^{-\frac{1}{2}}\mathcal{L}^{\frac{1}{2}}
\end{equation}
We will find the parameter space for which the thermal production rate 
becomes larger than the expansion rate $H$, i.e. $\Gamma /H>1$, within this temperature range  but outside the resonances, which we take as condition for thermalization. We thus compute $\Gamma/H$ when $2\epsilon T V_D /m_s^2 = x> 1$, where $x$ is a positive real parameter, namely when
\begin{equation}
\label{Tfo}
T= 0.596 \textrm{  MeV}x^{\frac{1}{4}}\epsilon^{-\frac{1}{4}}\left({m_s}/{\textrm{eV}}\right)^\frac{1}{2}\mathcal{L}^{-\frac{1}{4}}~.
\end{equation}
Since we are computing $\Gamma/H$ always at this temperature, when $\Gamma/H$=1 this is also the freeze-out temperature $T_{\rm f.o.}$.
We choose $x=3$ for Figs.~\ref{fig:allreslim2}  to~\ref{fig:appresliml5}. For this evaluation, we neglect the 1 and the $V_T$ terms in the denominator of the matter mixing angle, Eq.~\eqref{eq:mattermixing}. In the Std cosmology, the
$\Gamma /H>1$ thermalization condition is fulfilled and thus rejects the mixings
\begin{equation}
\label{eq:stdtherm}
    \sin^2(2\theta) > 119 \frac{{(1-x)^2}}{x^\frac{3}{4}} \left(\frac{\epsilon}{3.15}\right)^{\frac{3}{4}}\left(\frac{m_s}{\textrm{eV}}\right)^{-\frac{3}{2}}\mathcal{L}^{\frac{3}{4}}\left(\frac{g}{10.75}\right)^{\frac{1}{2}}~,
\end{equation}
and for ST1, it rejects
\begin{equation}
\label{eq:st1therm}
    \sin^2(2\theta) > 6.17\times10^8\frac{{(1-x)^2}}{x} \left(\frac{\epsilon}{3.15}\right)\left(\frac{m_s}{\textrm{eV}}\right)^{-2}\mathcal{L}\left(\frac{g}{10.75}\right)^{\frac{1}{2}}~.
\end{equation}
Results for K and ST2 are given in Appendix \ref{ssecapp:thermalization}. The region where thermalization is reached is shown in each panel of Figs.~\ref{fig:allreslim2}  to~\ref{fig:appresliml5} shaded  in cyan with the label ``therm.". When the freeze-out temperature $T_{\rm f.o.}$ in Eq.~\eqref{Tfo} becomes smaller than 5 MeV the ``therm." region for non-standard cosmologies should become the equal to the standard cosmology region (since sterile neutrino would be in equilibrium after the cosmology becomes standard).  This means that in all our figures for non-standard cosmologies the  ``therm." cyan region  has a vertical left boundary at $T_{\rm f.o.}=5$ MeV (because the Std. region does not extend to the left of this boundary). 

 In the darker cyan region, sterile neutrinos would be as abundant or almost as abundant as active neutrinos, thus this regions is forbidden by the $\Delta N_\textrm{eff}<$ 0.4  BBN limit (see below). Due to the entropy increase of interacting species  between $T_{\rm f.o.}$  and the freeze-out  of active neutrinos at 3 MeV,  when $T_{\rm f.o.}$  becomes larger than about 250 MeV (above the QCD phase transition), the sterile neutrino number density is diluted by a factor of about 5,  so that sterile neutrinos become allowed by the $\Delta N_\textrm{eff}$  limit.  This transition is indicated with a lighter cyan shade. Another transition to a yet lighter cyan is indicated when $T_{\rm f.o.}$ becomes larger than 1 GeV and the entropy dilution factor becomes close to 8.

The bound in Eq.~\eqref{eq:st1therm} is valid as long as the conditions in Eq.~\eqref{eq:thermtemprange} are satisfied which can only happen if the lepton number is large enough, namely for
\begin{equation}
\label{eq:Llimit}
\mathcal{L} > 2.1 \times 10^{-6}\left(\frac{\epsilon}{3.15}\right)^{\frac{1}{3}}\left(\frac{m_s}{\textrm{eV}}\right)^{\frac{2}{3}} \left(\frac{x}{3}\right)^{\frac{1}{3}}~.
\end{equation}
This condition agrees with Refs.~\cite{Hannestad:2012ky, Saviano:2013ktj}  (assuming $m_s$ is much larger than the active neutrino mass, as we do here) in that for $m_s \simeq$ eV,  the mass range required for MiniBoone/LSND,  enough suppression of the mixing in matter by the $V_D$ term to avoid thermalization requires a lepton number $\mathcal{L} > 10^{-5}$ (so the effect of $\mathcal{L}$  less than approximately 10$^{-5}$ is similar to that of $\mathcal{L}=0$~\cite{Hannestad:2012ky}). As can be seen in  Fig.~\ref{fig:appresliml5} in the  MiniBoone/LSND parameter region (hatched in black) in the standard cosmology (upper left panel) sterile neutrinos are partially or fully thermalized ($\Delta N_\textrm{eff} \approx 1$) for  a lepton asymmetry 
$\mathcal{L} = 10^{-5}$, and for this and smaller values of $\mathcal{L}$ would thus be in tension with CMB/BBN bounds on $N_\textrm{eff}$ (see Sec.~\ref{sec:limits}). Comparing our bound in Eq.~\eqref{eq:stdtherm} for the standard cosmology and $\mathcal{L} \simeq 10^{-2}$ to the $\Delta N_\textrm{eff} = 0.6$ limit in the upper panel of  Fig. 4 of Ref.~\cite{Hannestad:2012ky}, we see that both are similar in magnitude and shape.

The lower limit in Eq.\eqref{eq:Llimit} with $x=1$ is similar to the lower limit necessary to have a resonance $\mathcal{L}_{\textrm{reslim}}$  in Eq.~\eqref{eq:Lreslimit}, since the $V_D$ term needs to be equal to the sum of the other two terms in the denominator of Eq.~\eqref{eq:mattermixing}  to have a resonance and in our case the $V_T$ term is small. This is why in  Figs.~\ref{fig:allreslim2} to \ref{fig:appresliml5} the cyan thermalization region finishes at the boundary of the dark gray region of no resonance production.

So far we have considered the condition for chemical equilibrium of sterile neutrinos. We would like to point out that kinetic decoupling happens before chemical decoupling. The sterile neutrino scattering rate contains an extra $\sin^2 \theta$ factor over the production rate. Thus sterile neutrinos that are not in chemical equilibrium are also not in kinetic equilibrium.

\section{Constraints and regions of interest}
\label{sec:limits}

We consider here the same constraints and regions of interest in the mass-mixing plane for resonantly produced sterile neutrinos  in different pre-BBN cosmologies as we considered in our previous study of non-resonant production~\cite{Gelmini:2019esj,Gelmini:2019wfp}, modifying the relic density and characteristic momentum as necessary. The difference in the sterile neutrino production mechanism only affects the limits that depend on these quantities. We refer the reader to Ref.~\cite{Gelmini:2019wfp} for a more detailed discussion of the constraints that we briefly present, focusing here primarily on the changes due to the different production mechanism.
The resulting  limits and regions of interest are shown in  Figs.~\ref{fig:allreslim2} to Fig.~\ref{fig:appresliml5}, where we assume that the sterile neutrino only mixes with $\nu_e$. Since we assume fully resonant conversion (i.e. coherent and adiabatic) in computing the relic abundance and characteristic momentum, the limits that depend on the relic abundance and momentum are only shown within the red shaded wedges in which sterile neutrinos can be fully resonantly produced. We note that there are scalar-tensor cosmologies that produce wedges in between those we show for ST1 and ST2, always above the dot-dashed violet line.

Warm dark matter candidates are constrained by observations related to structure formation, which is suppressed below their free-streaming scales. In particular, measurements associated with $\sim 0.1-1$ Mpc scales~\cite{Baur:2017stq}, as probed by the Lyman-$\alpha$ forest absorption spectrum\footnote{Structure formation on similar scales can be also probed via DM halo counts~\cite{Boyarsky:2018tvu}.}, provide a strong limit on $\mathcal{O}$(keV) mass sterile neutrinos. Typically, bounds from Lyman-$\alpha$ are given in the literature in terms of the mass  $m_{\rm therm}$ of a thermally produced particle with a Fermi-Dirac spectrum. While for non-resonant production the average momentum of such particles $\langle \epsilon \rangle$ is close to 3 for all the cosmologies we consider, resonantly-produced sterile neutrinos have a lower average momentum that is close to or below 1 (see Eq.~\eqref{eq:reseps}). Following Ref.~\cite{Viel:2005qj} (see discussion in our previous study~\cite{Gelmini:2019wfp}), the limit on the sterile neutrino $m_s$ can be related to a given limit on the mass of a thermal relic $m_{\rm therm}$ as 
\begin{equation}
\label{eq:wdmdict-1}
m_s = 4.46 ~{\rm keV}~\left(\frac{\langle\epsilon\rangle}{3.15}\right)~
\left(\frac{m_\textrm{therm}}{\textrm{keV}}\right)^{\frac{4}{3}}~
\left(\dfrac{T_{\nu_s}}{T_{\nu_\alpha}}\right)~\left(\dfrac{0.12}{f_{\rm s,DM}~\Omega_{DM}~ h^2}\right)^{\frac{1}{3}}~,
\end{equation}
where  $T_{\nu_s}/T_{\nu_\alpha} = (10.75/g_{\ast})^{1/3}$. Taking $g_{\ast}=10.75$, we use Eq.~\eqref{eq:wdmdict-1} to translate to $m_s$ the limits of Ref.~\cite{Baur:2017stq}. We take the 2-$\sigma$ bound on warm dark matter from SDSS+XQ+HR data shown on the right panel of Fig. 6 of Ref.~\cite{Baur:2017stq}, but extend the limit horizontally to smaller masses, at 8\%  of the DM energy density. To obtain the corresponding Lyman-$\alpha$ bounds for sterile neutrinos,  which reject the gray vertical bands shown within the wedges in Fig.~\ref{fig:allreslim2},  we bounded the mass-mixing region by finding where the predicted DM fraction in each cosmology and for each $\mathcal{L}$ value becomes larger than the published 2-$\sigma$ limit.  Fully resonant sterile production affects the limits on $m_s$ found through Eq.~\eqref{eq:wdmdict-1}, due to the different values of $\langle \epsilon \rangle$ it predicts, which decrease as $\mathcal{L}$ decreases. 
For small $\mathcal{L}$, the corresponding $\langle \epsilon \rangle$ is also small, as shown in Eq. \eqref{eq:reseps}, i.e. a colder momentum distribution. Due to the diminished number density in Eq. \eqref{eq:sfoverdensity}, along with the $\left<\epsilon\right>$ dependence in Eq. \eqref{eq:wdmdict-1},  there are no  Ly-$\alpha$ bounds for $\mathcal{L}< 10^{-3}$,  thus the limit is only shown in Fig.~\ref{fig:allreslim2}.

Astrophysical indirect detection searches for X-ray emission probe the radiative decay $\nu_s\rightarrow\nu_\alpha\gamma$, whose lifetime $\tau_\gamma$  is~\cite{Shrock:1974nd, Pal:1981rm} 
\begin{equation}
\tau_\gamma^{-1} \simeq 1.38\times10^{-32} s^{-1} \Big(\dfrac{\sin^2(2\theta)}{10^{-10}}\Big)\Big(\dfrac{m_s}{\textrm{keV}}\Big)^5~,
\end{equation}
of sterile neutrinos in galaxies and galaxy clusters provide stringent constraints for $m_s > 1$ 
keV~\cite{Ng:2019gch,Perez:2016tcq,Neronov:2016wdd}.  
Additionally, observations of the diffuse extragalactic background radiation (DEBRA)~\cite{Boyarsky:2005us} limit
the integrated flux from sterile neutrino decays occurring between the recombination time $t_{\rm rec}$ and the present (thus it extends to $\tau= t_{\rm rec}$).   However, these limits scale with the density of sterile neutrinos and affect neutrinos with masses $m_s \geq 1 \textrm{ keV}$. Since our regions of interest are either already bounded by other limits at the keV scale or do not reach this mass, we do not show the X-ray/DEBRA bounds.

BBN imposes  a limit on $N_\textrm{eff}$, the effective number of active neutrino species, contributing to the energy density during BBN. In addition to $N_{\rm eff} = 3.045$ coming from active neutrinos~\cite{Mangano:2005cc, deSalas:2016ztq}, fully resonantly produced sterile neutrinos provide an extra contribution of
\begin{equation} 
\label{eq:neffeq}
\Delta N_{\rm eff} = N_{\rm eff} - 3.045 \simeq  \Big(\dfrac{\langle \epsilon \rangle}{3.15}\Big) \Big(\dfrac{n_{\nu_s}}{n_{\nu_{\alpha}}}\Big)\Big(\frac{10.75}{g_\ast}\Big)^{1/3}.
\end{equation}
Using Eq.~\eqref{eq:SFproduction}, the current  $95\%$ confidence level BBN bound\footnote{This bound is similar in magnitude to the $N_{\rm eff}$ constraint from CMB observations of Planck-2018~\cite{Aghanim:2018eyx}, which is however only applicable to sterile neutrinos with mass of $m_s \ll 1$ eV. Bounds on effective sterile neutrino mass $m_{s, eff}$ or the sum of active neutrino masses~\cite{Aghanim:2018eyx,Choudhury:2018sbz} from Planck-2018, BICEP2/Keck and BAO data do not significantly affect our results.} of $N_\textrm{eff}<3.4$~\cite{Tanabashi:2018oca} can be thus translated into an upper bound on the lepton number
\begin{equation}
\label{resneff}
    \mathcal{L} = 0.36 ~ \Delta N_\textrm{eff} \left(\frac{\left<\epsilon\right>}{3.15}\right)^{-1}\left(\frac{g_{\nu_s}}{10.75}\right)^\frac{1}{3} < 0.14 \left(\frac{\left<\epsilon\right>}{3.15}\right)^{-1}\left(\frac{g_{\nu_s}}{10.75}\right)^\frac{1}{3}~.
\end{equation}
Since the average momentum of fully resonantly produced sterile neutrinos is  $\left<\epsilon\right><3.15$ (see Eq.~\eqref{eq:reseps}), for $\Delta N_\textrm{eff}< 0.4$ the values of $\mathcal{L}$ chosen for Figs.~\ref{fig:allreslim2} to \ref{fig:appresliml5} easily evade this limit in the absence of thermalization. In the dark cyan shaded region where thermalization occurs, sterile neutrinos have the same (or close to the same) relic number density as one specifies of active neutrinos, i.e. $\Delta N_\textrm{eff} = 1$ (or close to 1, depending on entropy dilution) which is forbidden by the present limit (in the lighter cyan regions entropy dilution brings the sterile neutrino number density to be below the present limit, as explained above).  

Sterile neutrinos produced in supernova explosions could result in significant energy loss \cite{Kainulainen:1990bn}, disfavoring the brown  region of parameter space labelled ``SN" displayed in Figs.~\ref{fig:allreslim2} and \ref{fig:allreslim1}. It is difficult to exclude this region entirely, due to large uncertainties associated with the process~\cite{Abazajian:2001nj} (for recent related studies of sterile neutrinos mixing with $\nu_{\mu}$ and $\nu_{\tau}$ see Ref.~\cite{Arguelles:2016uwb,Raffelt:2011nc}).

Laboratory experiments can directly probe sterile neutrinos in the eV and keV mass-ranges by searching for active neutrino appearance or disappearance due to active-sterile neutrino mixing. These searches are completely independent of cosmology. In the eV-mass range, we display in  Figs.~\ref{fig:allreslim2} and \ref{fig:allreslim1} the combined limits (denoted by ``R'', green shading) from Daya Bay~\cite{An:2016luf}, Bugey-3~\cite{Declais:1994su} and PROSPECT~\cite{Ashenfelter:2018iov}, assuming sterile mixing with $\nu_e$. We highlight the regions (densely
hatched in black) corresponding to the anomalous signals\footnote{We stress that these signals are in strong tension with results from IceCube~\cite{TheIceCube:2016oqi} and MINOS~\cite{Adamson:2017uda}.} in LSND~\cite{Aguilar:2001ty} and MiniBooNE~\cite{Aguilar-Arevalo:2013pmq,Aguilar-Arevalo:2018gpe} data, 
reproduced from Fig.~4 of Ref.~\cite{Aguilar-Arevalo:2018gpe}. Additionally, we display (3 black vertical elliptical contours) the anomalous signal regions from DANSS~\cite{Alekseev:2018efk} and NEOS~\cite{Ko:2016owz} data, following Fig. 4 of Ref.~\cite{Gariazzo:2018mwd}. The eV-mass parameter space region will be further tested with KATRIN~\cite{megas:thesis} (solid blue lines) and PTOLEMY~\cite{Betti:2019ouf} (solid orange lines) experiments. In the keV-mass range, KATRIN~\cite{Wolf:2008hf,Mertens:2015ila} and its upgraded version TRISTAN~\cite{Mertens:2018vuu} (solid blue lines), as well as the upcoming HUNTER~\cite{Smith:2016vku} (solid magenta lines) experiment, will be able to probe a significant portion of the active-sterile mixing parameter space\footnote{The effects of lepton asymmetry on cosmological sterile neutrino bounds at keV mass range have been also recently discussed in Ref.~\cite{Benso:2019jog}.}.

Limits on neutrinoless double-beta decay constrain sterile neutrinos mixed with $\nu_e$, if neutrinos are Majorana fermions, through their contribution $\left<m\right>_s = m_s \sin^2(\theta) e^{i\beta_s}$ to the effective electron neutrino Majorana mass ($\beta_s$ is a Majorana CP-violating phase).  The present bound on the magnitude of this effective mass is $|\langle m\rangle|<0.165$ eV~\cite{KamLAND-Zen:2016pfg}. Hence, this implies an  upper limit (denoted ``$0\nu\beta\beta$", solid orange line) of $m_s\sin^2(2\theta)<0.660$ eV. We note that this bound is not completely robust. The contribution of the sterile neutrino might interfere with the contributions from the active ones, leading to a suppression in the effective Majorana mass and, therefore, avoiding the experimental bounds~\cite{Abada:2018qok}.    

\section{Summary of main results}
\label{sec:mainresults}

We studied the cosmological dependence of resonant sterile neutrino production for sterile neutrino masses between $10^{-2}~{\rm eV} < m_s < 10^3~{\rm keV}$, assuming different pre-BBN cosmologies before the temperature of the Universe was 5 MeV, which all transition to the standard cosmology for $T\leq$ 5 MeV. We derived general expressions for relevant quantities using a simple parametrization of the expansion of the Universe that covers a broad class of cosmologies and which has, as special cases, the particular cosmologies we studied as examples: the standard cosmology (Std), kination (K), and two scalar-tensor models denoted ST1 and ST2. Resonant production can only happen in the presence of a large lepton asymmetry.

We show the regions of interest and different astrophysical and cosmological limits for a $\nu_s$ mixed only with $\nu_e$, appropriate for our scenarios in  Figs.~\ref{fig:allreslim2} to ~\ref{fig:appresliml5} for $\mathcal{L} = 10^{-2}, 10^{-3}, 10^{-4}$ and $10^{-5}$, respectively, 
for the four example cosmologies.

The vertical solid black line labeled  
``no res. prod." shows the maximum value of the mass in each panel for the given $\mathcal{L}$ and $\epsilon = p/T = 1$ for which resonant production can take place. No resonance production can occur in dark gray shaded region to the right of this line.  In all the cosmologies we studied the resulting characteristic momentum is  $\epsilon < 1$.  The vertical strip to the left of this limit, diagonally hatched  in red and labeled ``two pop.'', indicates the approximate range in which non-resonant production may take place after resonant production. Within this mass range the total momentum distribution would consist of an overlap of two populations of the same sterile neutrino with different characteristic momenta: a colder resonantly produced one with $\epsilon < 1$ and a hotter one with larger characteristic momentum  $\epsilon \simeq 3$ (in comparison, the characteristic momentum for non-resonant production we found in Refs.~\cite{Gelmini:2019wfp, Gelmini:2019esj} is $\sim 3$ for all the considered cosmologies).  We obtained this range by requiring the temperature of maximum non-resonant production $T_{\rm max}$ to be smaller than the resonance temperature. Since even for temperatures lower than $T_{\rm max}$ there can still be a considerable rate of non-thermal production, the lower limit of this narrow band is not strict. We note that within this band, the limits on non-resonant production presented e.g. in Refs.~\cite{Gelmini:2019wfp,  Gelmini:2019esj} would apply, which are not shown in the figures of this paper (including thermalization limits).

We identified the region (above the dot-dashed violet diagonal lines in each panel in  Figs.~\ref{fig:allreslim2} to ~\ref{fig:appresliml5})   in the mass-mixing plane where both adiabaticity and coherence can happen simultaneously. In this region the damping term appearing in the Boltzmann equation is always negligible  (below the dot-dashed violet diagonal lines the damping term is important at resonance).  Further, we specifically identified where in the different cosmologies there is fully resonant conversion (in contrast to resonant scattering-induced production), which requires that two conditions hold during resonance: coherence and adiabaticity. A fully resonant conversion leads to a very effective transfer of the initial active neutrino lepton number into a population of sterile neutrinos with number density   $n_{\nu_s, {\rm res}} \simeq \Big(\sum L_{\nu_{\alpha}}\Big) n_{\gamma}$, independent of the mixing angle,   and  a characteristic momentum  $\epsilon <1$ that diminishes with the initial lepton number (see Eq.~\eqref{eq:reseps}). Fully resonant conversion in each of the four cosmologies that we consider occurs in the red shaded wedges in  Figs.~\ref{fig:allreslim2} to ~\ref{fig:appresliml5}, delimited from above by the coherence limit and by the adiabaticity limit from below.  We clearly see in the figures that these wedges change considerably with the cosmology, allowing sterile neutrinos that in the standard cosmology would not be fully resonantly produced to be produced in this manner in a non-standard one. This type of production could in principle be distinguished by measuring the relic density and spectrum.  To the left of the  vertical line labeled $T_{\rm res}= 5$ MeV, resonant production would happen at temperatures smaller than 5 MeV where all our cosmologies become identical to the standard one, thus only the red wedge of the standard cosmology extends to the left of this line.

We note that practically all fully resonantly produced sterile neutrinos would constitute a part of the hot dark matter (except possibly for the ST1 and K cosmologies and $\mathcal{L}= 10^{-3}$, in which case they might be a subdominant warm dark matter component with mass close to a keV). Ly-$\alpha$-HDM limits reject  the gray region shown within the wedges in the plots for $\mathcal{L}= 10^{-2}$.
To the left of this  Ly-$\alpha$-HDM rejected mass range for $\mathcal{L}= 10^{-2}$, 
and  to the left of the dark gray ``no res. prod. region" (where resonant production can happen) for $\mathcal{L}= 10^{-3}, 10^{-4}$ and 10$^{-5}$ fully resonant production yields a viable dark matter component. The main distinguishing characteristic of these sterile neutrinos would be their colder spectrum (with respect to non-resonantly produced neutrinos). Sterile neutrinos with a colder spectrum could potentially be detected in the future by their impact on the matter power spectrum $P(k)$. In particular, sterile neutrinos would lead to a suppression of $P(k)$ for $k > k_{\rm nr}$ above the scale associated with $k_{\rm nr}$ (smaller length scales), where they become non-relativistic,  proportional to their density fraction 
$f_{s, DM}=\Omega_s/ \Omega_{\rm DM}$. This effect would be in addition to the qualitatively  similar effect expected from the active neutrino bath~$P(k)$~\cite{Lesgourgues:2006nd, Abazajian:2013oma,Abazajian:2016yjj}. 

We showed that the light sterile neutrinos that could be detected in reactor neutrino experiments or the KATRIN and PTOLEMY experiments could be fully resonantly produced in the early Universe in several of the cosmologies we consider.

\section{Conclusions}
\label{sec:summary} 

Production through resonant active-sterile flavor oscillations constitutes a major early Universe production mechanism that is often considered for sterile neutrinos. We studied the cosmological evolution dependence of sterile neutrino resonant production, considering several different pre-BBN cosmologies before the temperature of the Universe was 5 MeV. This study complements our previous work on production via  non-resonant active-sterile neutrino flavor oscillations  in the same pre-BBN cosmologies~\cite{Gelmini:2019esj,Gelmini:2019wfp} and further explores the extent to which sterile neutrinos can act as sensitive probes of early Universe cosmology before the BBN, which has not yet been tested. 

Resonant sterile neutrino production requires a large lepton asymmetry in the active neutrino background, much larger than the baryon asymmetry, whose origin we do not discuss. Assuming that the sterile neutrino mixes only with one active neutrino (which for the figures we take to be $\nu_e$), for each value of the lepton number $\mathcal{L}$ we pointed out the maximum sterile neutrino mass value for which resonant production can occur. We also showed the approximate range of masses just below this limit where non-resonant production can take place after resonant production (i.e. in which the  maximum temperature of non-resonant production is below the resonance temperature). In this case, two populations of sterile neutrinos would contribute to the total sterile neutrino spectrum: one colder, due to resonant production (with characteristic momentum $p < T$) and another hotter (with characteristic momentum $p \simeq 3 T$). Two populations of sterile neutrinos have been previously found in the literature, but to the best of our knowledge the approximate range of masses for which this can happen has not been discussed.

We determined the region in sterile neutrino mass-mixing parameter space where adiabaticity and coherence are simultaneously possible (and in which quantum damping is  negligible even at the resonance). We identified ``wedges" in the parameter space for each cosmology we considered where coherence and adiabaticity hold simultaneously, thus yielding a fully resonant active-sterile conversion (thus a number density  equal to  the active neutrino lepton asymmetry  times the photon number density, independently of the mixing angle). Different locations of these regions in the mass mixing plane for different cosmologies demonstrates that sterile neutrinos are sensitive to the assumed early Universe history and can have varying momentum spectra and density for a given same mass and active-sterile mixing.

Finally, we found that the cosmological bounds on eV mass-scale sterile neutrinos are relaxed for fully resonantly produced sterile neutrinos in several of the cosmologies we considered, including the standard one for a lepton asymmetry larger than $\sim 10^{-5}$.  This  allows for the results reported from the LSND, and MiniBooNE short-baseline as well as the DANSS and NEOS reactor neutrino experiments to be unrestricted by cosmology. More so, sterile neutrinos in this mass-range are also within the reach of the KATRIN and PTOLEMY experiments.

\acknowledgments
\addcontentsline{toc}{section}{Acknowledgments}
 
The work of G.B.G., P.L. and V.T. was supported in part by the U.S. Department of Energy (DOE) Grant No. DE-SC0009937.

\appendix

\section{Additional formulas}
\label{sec:appendixformulas}

\subsection{Temperature of maximum non-resonant production}
\label{ssecapp:maxtemp}

We recomputed for the different cosmologies we considered the temperature  $T_{\rm max}$ at which the non-resonant production rate integrated over momentum, denoted $\gamma$ in the DW paper~\cite{Dodelson:1993je}, over $H$
\begin{equation}
\label{eq:Gamma-over-H}
\frac{\gamma}{H}=\frac{d}{d \ln{T}}\left(\frac{n_{\nu_s}}{n_{\nu_\alpha}}\right)~,
\end{equation}
is maximum,  as done for the standard cosmology in Ref.~\cite{Dodelson:1993je} (see Eqs. (4) to (6) of Ref.~\cite{Dodelson:1993je} and also \cite{Kainulainen:1990ds}). Note that in our previous study of Ref.~\cite{Gelmini:2019wfp}, we refer to $T_{\rm max}$ as the maximum of $(\Gamma f_{\nu_\alpha}/HT)$, not integrated over momenta, which thus depends on $\epsilon$. We recomputed $T_\textrm{max}$ for each cosmologies by numerically plotting 
$(\gamma/H)$ and changing the $x$ factor in front of the integral in Eq. (6) of Ref.~\cite{Dodelson:1993je} to $x^{1-\beta/3}$.
Our value for the standard cosmology differs from that of DW because we use $B=10.88\times10^{-9} \textrm{ GeV}^{-4}$, which is the right value for the $\nu_e \leftrightarrow \nu_{s}$ transitions considered in our paper, while DW used $B\approx3\times 10^{-9} \textrm{ GeV}^{-4}$, as is appropriate for $\nu_{\mu}$ or $\nu_{\tau}$ to $\nu_s$ transitions at temperatures at which $\mu$ and $\tau$ charged leptons are not in equilibrium (see Eq.\eqref{eq:bprefac}). Our result for the standard cosmology coincides with the $T_{\rm max}$ given in Ref.~\cite{Kainulainen:1990ds},
\begin{equation}
\label{eq:Stdmax}
    T_{\textrm{max}}^\textrm{Std} = T_{\textrm{max}}^\textrm{ST2} \simeq 108~\textrm{MeV} \left(\frac{m_s}{\textrm{keV}}\right)^{\frac{1}{3}}~.
\end{equation}
For the standard and ST2 cosmologies, $\beta = 0$.
For ST1 ($\beta = -0.8$),
\begin{equation}\label{eq:st1tmax}
    T_{\textrm{max}}^{\rm ST1} \simeq 118~\textrm{MeV} \left(\frac{m_s}{\textrm{keV}}\right)^{\frac{1}{3}}~,
\end{equation}
and for kination ($\beta = 1$)
\begin{equation} \label{eq:ktmax}
    T_{\textrm{max}}^{\rm K} \simeq 94.5~\textrm{MeV} \left(\frac{m_s}{\textrm{keV}}\right)^{\frac{1}{3}}~.
\end{equation}

\subsection{Combined resonant and non-resonant production}
\label{ssecapp:resvnonres}

In Section~\ref{ssec:resvsnonres},  we derived a lower limit on the sterile neutrino mass for non-resonant production to happen after resonant production.
In the K cosmology ($\eta = 1$, $\beta = 1$) this is
\begin{equation}
\label{eq:kinresmax}
    m_s > m_\textrm{non-res}^{\rm K} = 1.61\times10^{4}\textrm{ keV}~\epsilon_\textrm{res}^{\frac{3}{2}}\mathcal{L}^{\frac{3}{2}}~,
\end{equation}
and for ST2 ($\eta = 0.03$, $\beta = 0$)
\begin{equation}
\label{eq:st2resmax}
    m_s > m_\textrm{non-res}^{\rm ST2} = 3.59\times10^{4}\textrm{ keV}~\epsilon_\textrm{res}^{\frac{3}{2}}\mathcal{L}^{\frac{3}{2}}~.
\end{equation}

\subsection{Coherence}
\label{ssecapp:coherence}

The coherence condition in Eq.~\eqref{eq:coherence}, $\Gamma_{\alpha}^{-1} > l_{\rm res}$, given as a function of the $\eta$ and $\beta$ parameters in Eq.~\eqref{eq:hnStd} is
\begin{equation}
\label{eq:gencoherence}
    \sin^2 (2\theta) <   2.19\times10^{-6}~(3.76)^{2\beta}~\eta^2\epsilon^{-\frac{1}{2}-\frac{\beta}{2}}\left(\frac{m_s}{\textrm{keV}}\right)^{-3+\beta}\mathcal{L}^{\frac{3}{2}-\frac{\beta}{2}}\left(\frac{T_\textrm{tr}}{5\textrm{ MeV}}\right)^{-2\beta}\left(\frac{g_{\ast}}{10.75}\right)~.
\end{equation}
For the K cosmology ($\eta = 1$, $\beta = 1$), this is
\begin{equation}
\label{eq:kincoherence}
    \sin^2 (2\theta)^{\rm K} < 3.10\times10^{-5} \epsilon^{-1}\left(\frac{m_s}{\textrm{keV}}\right)^{-2}\mathcal{L}\left(\frac{T_\textrm{tr}}{5\textrm{ MeV}}\right)^{-2}\left(\frac{g_{\ast}}{10.75}\right)~,
\end{equation}
while for the ST2 cosmology ($\eta = 0.03$, $\beta = 0$) this is
\begin{equation}
\label{eq:st2coherence}
    \sin^2 (2\theta)^{\rm ST2} < 2.24\times10^{-9} \epsilon^{-\frac{1}{2}}\left(\frac{m_s}{\textrm{keV}}\right)^{-3}\mathcal{L}^{\frac{3}{2}}\left(\frac{g_{\ast}}{10.75}\right)~.
\end{equation} 

\subsection{Adiabaticity}
\label{ssecapp:adiabaticity}
The adiabaticity condition $\gamma > \gamma_\textrm{lim} \geq 1$ of Eq.~\eqref{eq:adiab} in terms of the $\eta$ and $\beta$ parametrs of $H$ in Eq.~\eqref{eq:hnStd} is
\begin{equation}
\label{eq:genadiabaticity}
    \sin^2 (2\theta) > 2.34\times10^{-11}(3.76)^\beta~ \eta~\gamma_\textrm{lim}~\epsilon^{\frac{1}{4}-\frac{\beta}{4}}\left(\frac{m_s}{\textrm{keV}}\right)^{\frac{\beta}{2}-\frac{1}{2}}\mathcal{L}^{-\frac{3}{4}-\frac{\beta}{4}}\left(\frac{T_\textrm{tr}}{5\textrm{ MeV}}\right)^{-\beta}\left(\frac{g_{\ast}}{10.75}\right)^\frac{1}{2}~.
\end{equation}
For K cosmology ($\eta = 1$, $\beta = 1$) this becomes
\begin{equation}
\label{eq:kinadiabaticity}
    \sin^2(2\theta)^{\rm K}>8.80\times10^{-11}\gamma_\textrm{lim}~\mathcal{L}^{-1}\left(\frac{T_\textrm{tr}}{5\textrm{ MeV}}\right)^{-1}\left(\frac{g_{\ast}}{10.75}\right)^{\frac{1}{2}}~,
\end{equation}
and for ST2 ($\eta = 0.03$, $\beta = 0$) cosmology it is 
\begin{equation}
\label{eq:st2adiabaticity}
    \sin^2(2\theta)^{\rm ST2}>7.48\times10^{-13}\gamma_\textrm{lim}~\epsilon^\frac{1}{4}\left(\frac{m_s}{\textrm{keV}}\right)^{-\frac{1}{2}}~\mathcal{L}^{-\frac{3}{4}}\left(\frac{g_{\ast}}{10.75}\right)^{\frac{1}{2}}~.
\end{equation}

\subsection{Thermalization}
\label{ssecapp:thermalization}
The thermalization limit given as a function of the $\eta$ and $\beta$ parameters in Eq.~\eqref{eq:hnStd} rejects
\begin{equation}
\label{eq:gentherm}
    \sin^2(2\theta) > 119\frac{{(1-x)^2}}{x^\frac{3-\beta}{4}} (0.09)^\beta \eta \left(\frac{\epsilon}{3.15}\right)^{\frac{3-\beta}{4}}\left(\frac{m_s}{\textrm{eV}}\right)^{-\frac{3-\beta}{2}}\mathcal{L}^{\frac{3-\beta}{4}}\left(\frac{g}{10.75}\right)^{\frac{1}{2}}~,
\end{equation}
For the K cosmology ($\eta = 1$,$\beta = 1$), the limit rejects
\begin{equation}
\label{eq:ktherm}
     \sin^2(2\theta) > 10.5\frac{{(1-x)^2}}{x^\frac{1}{2}} \left(\frac{\epsilon}{3.15}\right)^{\frac{1}{2}}\left(\frac{m_s}{\textrm{eV}}\right)^{-1}\mathcal{L}^{\frac{1}{2}}\left(\frac{g}{10.75}\right)^{\frac{1}{2}}~,   
\end{equation}
while for the ST2 cosmology ($\eta = 0.03$, $\beta = 0$) this is
\begin{equation}
\label{eq:st2therm}
    \sin^2(2\theta) > 3.56 \frac{{(1-x)^2}}{x^\frac{3}{4}}\left(\frac{\epsilon}{3.15}\right)^{\frac{3}{4}}\left(\frac{m_s}{\textrm{eV}}\right)^{-\frac{3}{2}}\mathcal{L}^{\frac{3}{4}}\left(\frac{g}{10.75}\right)^{\frac{1}{2}}~,
\end{equation}

\clearpage
\bibliography{sternumodcos}
\addcontentsline{toc}{section}{Bibliography}
\bibliographystyle{JHEP}
\end{document}